\documentclass[%
 aip,
 amsmath,amssymb,
 reprint,%
author-year,%
]{revtex4-1}

\usepackage{graphicx}
\usepackage{dcolumn}
\usepackage{bm}
\usepackage[mathlines]{lineno}

\usepackage[utf8]{inputenc}
\usepackage[T1]{fontenc}
\usepackage{mathptmx}
\usepackage{etoolbox}
\usepackage{xcolor}

\makeatletter
\def\@email#1#2{%
 \endgroup
 \patchcmd{\titleblock@produce}
  {\frontmatter@RRAPformat}
  {\frontmatter@RRAPformat{\produce@RRAP{*#1\href{mailto:#2}{#2}}}\frontmatter@RRAPformat}
  {}{}
}%
\makeatother
\begin{document}

\preprint{AIP/123-QED}

\title[]{Applications of a moving surface drag model: From Stokes waves to offshore wind farms}

\author{M. Ayala}
\email{mayala5@jhu.edu}
\affiliation{Department of Mechanical Engineering, Johns Hopkins University, Baltimore, MD, USA}

\author{Z. Sadek}
\affiliation{Mechanical and Materials Engineering, Portland State University, Portland, OR, USA}

\author{R.B. Cal}
\affiliation{Mechanical and Materials Engineering, Portland State University, Portland, OR, USA}

\author{D.F. Gayme}
\affiliation{Department of Mechanical Engineering, Johns Hopkins University, Baltimore, MD, USA}

\author{C. Meneveau}
\affiliation{Department of Mechanical Engineering, Johns Hopkins University, Baltimore, MD, USA}
\date{\today}

\begin{abstract}
Offshore wind farms operate within the marine atmospheric boundary layer, where ocean waves can modify air-sea momentum transfer, turbine inflow, and wake recovery. Wave-resolved simulations can capture phase-dependent wind--wave interactions, but their computational cost and implementation complexity limit their use in large atmospheric and wind-farm domains. This study describes applications and tests of the recently proposed moving surface drag (MOSD) model, a flat-bottom wall-stress formulation for horizontally resolved moving waves, to a sequence of increasingly complex wind--wave and offshore wind-energy applications. The model is first evaluated for turbulent flow over Stokes-like waves using laboratory measurements, where it reproduces mean velocity profiles and captures the qualitative structure of wave-induced motions, although their amplitude is underpredicted. MOSD is then applied to broadband wave spectra fields and offshore wind-farm boundary layers, showing good agreement with  mean velocity profiles from wave-resolved simulations across different wave ages, grid resolutions, and turbine spacings. Finally, exploratory offshore wind-farm simulations using MOSD under conventionally neutral conditions demonstrate its application at wind-farm scale. Large-scale waves leave detectable signatures in near-surface velocity spectra, but these signatures decay with height and do not produce clear peaks in temporal spectra of aggregate turbine power. Overall, the results demonstrate that the MOSD model provides a tractable framework for incorporating resolved wave effects in wind-wave and offshore wind-farm simulations. 
\end{abstract}

\maketitle

\section{\label{sec:intro}Introduction}

Air-sea momentum exchange plays a central role in the dynamics of the marine atmospheric boundary layer (MABL) and directly affects weather prediction, climate modeling, marine operations, and offshore wind-energy applications in particular.  Unlike boundary layers over land, the MABL develops over a moving and deforming lower boundary. Ocean waves influence the airflow through form drag, viscous and tangential stress, and wave-induced momentum fluxes, all of which depend on wave geometry, wave age, steepness, propagation direction, and the relative motion between the waves and the wind \citep{sullivan_annual2010,ayet2022}. 

Experiments and numerical simulations have shown that waves can modify mean velocity profiles, effective aerodynamic roughness, turbulent momentum fluxes, and Reynolds stresses \citep{edson2013,grare2018,yousefi_2020}. Beyond these effects on mean and turbulence statistics, the influence of waves also manifests through motions that are systematically correlated with the wave phase. These wave-coherent motions, in which the velocity and stress fields vary with wave phase, have been observed in laboratory measurements, field studies, and wave-resolved simulations \citep{sullivan_2000,Hristov2003,buckley2016}. Thus, ocean waves do not simply act as an equivalent roughness; they can also organize near-surface turbulence and redistribute momentum in a phase-dependent manner.

These wave-induced changes have the potential to directly affect offshore wind energy harvesting, because turbines operate within the part of the atmosphere shaped by surface stress and turbulent momentum transport. Changes in the mean velocity profile affect the inflow across the rotor, while changes in turbulent momentum exchange influence how quickly turbine wakes recover downstream \citep{porteagel_review_2020,stevens_review_2017}. This connection is especially important in large wind farms, where downstream turbines often operate in partially recovered wakes and rely on the vertical replenishment of momentum from the flow above \citep{Hamilton2012}. Therefore, even when wave-coherent motions are strongest near the surface, their influence on surface stress and turbulent transport can affect the turbine-relevant region of the boundary layer.

Recent laboratory studies have begun to clarify how these processes appear in wind--wave--wake interactions. \citet{fercak_2022} measured the wake of a fixed-bottom offshore wind turbine over moving waves and showed that wave phase modulates both the position and recovery of the wake, producing wave-dependent wake pumping, wake lifting, and Reynolds-stress variations. Building on those measurements, \citet{Mouchref2024} used quadrant analysis to show that the Reynolds shear-stress events responsible for turbulent momentum transport also depend on wave phase. These results indicate that waves can influence not only the incoming MABL, but also the turbulent mechanisms that control wake recovery within offshore wind farms.

Several numerical studies have examined these effects by explicitly representing the moving ocean surface. \citet{yang2014large} performed large-eddy simulations (LES) of offshore wind farms over broadband waves using a wave-following coordinate transformation, in which the physical domain bounded by the evolving sea surface is mapped to a rectangular computational domain. They showed that wave state can have an appreciable effect on wind-farm performance, with fully developed waves producing higher power extraction than fetch-limited seas because of their faster wave propagation and lower sea-surface resistance. Similarly, \citet{xiao2019} used a coupled LES framework with a wave-following grid to study floating offshore wind turbines and showed that swell-induced pitch motion can modulate turbulence intensity, Reynolds stresses, and the phase dependence of power extraction. More recently, \citet{Ciri2024} used LES with a hybrid level-set/immersed-boundary method to track the air--water interface and study wave--wake interactions for different wave ages, finding that waves can modify near-surface turbulence kinetic energy and wake recovery. These studies demonstrate the importance of wave effects for offshore wind-energy applications, but also highlight a modeling challenge: phase-resolved representation of the moving sea surface typically requires grid deformation, wave-following mappings, or interface-tracking methods.

Although wave-resolved simulations can capture pressure variations over the wave surface and wave-coherent motions, their computational cost can become restrictive for broadband wave fields, atmospheric Reynolds numbers, and wind-farm domains containing many turbines \citep{Deskos2021}. A more economical and simpler approach is to represent the sea surface through an equivalent aerodynamic roughness length, as is commonly done in wall-modeled LES using Charnock-type or wave-age-dependent parameterizations \citep{Deskos2021,ayalaSWARL2026}. Such models are simple and computationally inexpensive, but they describe the sea surface only in a phase-averaged sense. As a result, they cannot explicitly represent the local form drag produced by resolved wave slopes or the wave-coherent velocity and stress variations observed in experiments and wave-resolved simulations. This motivates modeling approaches that retain the efficiency of wall-modeled LES while allowing waves to affect the flow in a phase-dependent manner.

Recent work has therefore sought reduced-order representations of wave effects for wall-modeled LES. \citet{Aiyer_2024} developed a dynamic wave-spectrum drag model in which horizontally resolved wave modes are represented through a modeled wave-drag force, and unresolved wave components are incorporated through a dynamically determined roughness length. Rather than representing the resolved wave field through a spectral superposition of modal drag contributions at discrete wavenumbers, \citet{ayala2024} introduced the moving surface drag (MOSD) model, formulated as a local surface-geometry-based model for the drag induced by horizontally resolved moving waves. The model was developed to retain phase-dependent wave effects in a flat-bottom LES framework and was originally derived and validated for monochromatic waves and laboratory-scale wind--wave--wake interactions. 

The present study applies the MOSD framework to more practical MABL and offshore wind-energy applications. We consider increasingly complex wave fields and wind-farm configurations, including nonlinear wave shapes, broadband spectra, and offshore wind-farm boundary layers. We also introduce a geometry-based reference height and account for unresolved wave scales through an equivalent roughness length. The goal is to assess whether MOSD-based wall-modeled LES can provide a computationally tractable approach for representing wave effects in offshore wind-energy flows that does not rely on fully wave-resolved simulations.

\section{\label{sec:mosd}Moving Surface Drag (MOSD) Model for Wind over Waves}

Salient features of the MOSD model for air-sea interactions developed by \cite{ayala2024} are here summarized. The MOSD model is a wall-stress boundary condition that accounts for the drag generated by moving surfaces, such as ocean waves. It has previously been validated for monochromatic sinusoidal waves and has since been used to introduce wave effects in simulations of offshore wind turbine wake dynamics \citep{yuan_2026}. It has also been implemented within the WRF-LES framework for phase-aware simulations of turbulent flow over moving waves \citep{kale2026}.

The modeled wall stress is written as $\tau^{\mathrm w}_{iz}=\tau^{\mathrm{w-p}}_{iz}+\tau^{\mathrm{w-u}}_{iz}$, where $i=1,2$. It consists of two contributions: a pressure-induced component (form drag), $\tau^{\mathrm{w-p}}_{iz}$, and a component associated with unresolved effects, $\tau^{\mathrm{w-u}}_{iz}$, including viscous stress and drag due to small-scale roughness elements that are not explicitly resolved. The pressure-stress contribution $\tau^{\mathrm{w-p}}_{iz}$ accounts for the drag associated with resolved wave surfaces. The underlying assumption is that in a reference frame moving with the local wave phase velocity ${\bf C}$, and incoming relative velocity ${\bf u}_\Delta-{\bf C}$, the local flow can be represented as potential flow over a ramp of slope angle $\alpha(x,y)$. Here, ${\bf u}_\Delta$ is the horizontal air velocity at a reference height $\Delta$ above the mean wave elevation.

The average pressure over such a ramp is proportional to the square of the relative velocity multiplied by $\alpha/(\pi+\alpha)$, and the resulting pressure contribution to the modeled stress can be written as
\begin{equation}
\tau^{\mathrm{w-p}}_{i3} = \frac{\alpha}{\pi+\alpha}\, |(\boldsymbol{u_\Delta-C}) \cdot \boldsymbol{\hat{n}}|^2\,|\boldsymbol{\nabla} \eta|\, \text{H}\left[(\boldsymbol{u_\Delta}-\boldsymbol{C})\cdot \boldsymbol{\nabla} \eta \right].
\label{eq:mosd_wpm}
\end{equation}
Here, $\eta(x,y,t)$ is the known surface elevation as a function of horizontal position $(x,y)$ and time $t$, and $\alpha(x,y)=\arctan(|\boldsymbol{\nabla}\eta|)$. The pressure rise is produced by the component of the relative velocity normal to the surface. Therefore, only the normal projection of the relative velocity, $({\bf u}_\Delta-{\bf C})\cdot \hat{\bf n}$, contributes to the stress, where ${\bf \hat n}={\bf \nabla}\eta/|{\bf \nabla}\eta|$ is the unit normal vector in the horizontal plane.

Moreover, $\boldsymbol{C}$ is the local phase velocity of the wave, that is, the horizontal speed of the local vertical projection of the surface. It can be computed from $\eta(x,y,t)$ as
\begin{equation}
C_i = - \frac{\partial \eta}{\partial t}\,\frac{\partial \eta}{\partial x_i}\,\frac{1}{|\boldsymbol{\nabla}\eta|^2}.
\label{eq:wavevel}
\end{equation}
Note that $C_i$ may diverge at points where $|\nabla \eta| \to 0$, since $|\nabla \eta|$ appears in the denominator of Eq.~\eqref{eq:wavevel}. However, at such points $\alpha=0$ and $\partial \eta/\partial x_i=0$, so no singularity arises when evaluating $\tau^{\mathrm{w-p}}$. For a simple monochromatic wave propagating in the $x$-direction with surface elevation $\eta=a\cos(kx-\omega t)$, Eq.~\eqref{eq:wavevel} reduces to the usual phase speed, $C_1=c=\omega/k$.

For numerical convenience, we clip the phase velocity $C_i$ when $|\nabla \eta|$ approaches zero in Eq.~\eqref{eq:wavevel}, using the upper bound $|{\bf C}|_{\rm max}=\sqrt{g/(0.25k_p)}$. The use of $0.25k_p$ ensures that the cutoff corresponds to waves substantially faster than the fastest expected waves, with a characteristic wavelength four times that of the peak wave. 

The model assumes that streamline displacement causes a pressure drop, or prevents pressure recovery, on the leeside of the wave, so that the pressure force there can be neglected. Accordingly, the Heaviside function $\text{H}(x)=\frac{1}{2}(x+|x|)/x$ is used to impose the pressure force only on the windward side of the wave. Although flow separation has been observed experimentally \citep{buckley2020,veron2007}, the assumption of displaced streamlines leading to negligible leeside pressure is a strong modeling assumption and may not apply to all wind--wave conditions, such as swell or waves propagating against the wind.

 It is important to note limitations of MOSD. Most severely, the pressure-stress model relies on the assumption that a horizontally discretized wavy surface can be approximated as a set of locally straight, unconnected ramps, with the flow impinging horizontally and independently on each ramp \cite{ayala2024}. In reality, however, the flow near wave crests may exhibit strong acceleration and substantial pressure reduction, even reaching negative values, which can significantly affect the drag \citep{Sullivan_2018}. Such behavior cannot be captured accurately by the purely local potential-flow model adopted here. A more realistic description of the near-surface flow, even within a potential-flow framework, would require nonlocal information, that could be obtained, for example, by solving partial differential equations using eigenfunction expansions. Such approaches are significantly more computationally expensive than the present form-drag model, which requires only simple local evaluations and area integration. 

The tangential viscous stress and unresolved surface form drag are modeled through a friction-factor parameterization:
\begin{equation}
\tau^{\mathrm{w-u}}_{iz} = \frac{1}{2}\, C_f(Re_\Delta, z_0^u)\,|\bm{u_{\Delta_{\rm m}}}| u_{\Delta_{\rm m},i}.
\label{eq:visc_stress}
\end{equation}
The friction factor $C_f$ is determined using the generalized Moody-diagram fit 
~\citep{meneveau2020}. It depends on $Re_\Delta=|\bm{u_{\Delta_{\rm m}}}|\Delta_{\rm m}/\nu$ and  the unresolved roughness length $z_0^u$, and is written as
\begin{equation}
C_f(Re_\Delta,z_0^u) = 2\Biggl\{ \left[\frac{1}{2}C_{fs}(Re_\Delta)\right]^3 + \left[\frac{1}{\kappa}\ln\left(\frac{\Delta_{\rm m}-\eta}{z_0^u}\right)\right]^{-6} \Biggr\}^{1/3}.
\label{eq:cftotal}
\end{equation}

Here, $\Delta_{\rm m}$ is the LES wall-model matching height, defined as the location above the surface at which the LES velocity is sampled to obtain $\bm{u_{\Delta_{\rm m}}}$ in  Eq.~\ref{eq:visc_stress}. Selecting the matching location sufficiently far from the wall has been shown to help reduce the log-layer mismatch commonly observed in wall-modeled LES \citep{KawaiLarsson}. $C_{fs}(Re_\Delta)$ is the smooth-surface friction coefficient, and $\kappa=0.4$ is the von Kármán constant. The smooth-surface coefficient $C_{fs}(Re_\Delta)$ is obtained from a fit to the numerical integration of the equilibrium wall-model equation \citep{meneveau2020}. In its simplest form,
$$
C_{fs}(Re_\Delta) = 0.0288\,Re_\Delta^{-1/5}\left(1+577\,Re_\Delta^{-6/5}\right)^{2/3},
$$
where $Re_\Delta=|\bm{u_{\Delta_{\rm m}}}|\Delta_{\rm m}/\nu=|\bm{u_{\Delta_{\rm m}}}|^+\Delta_{\rm m}^+$. The second term in Eq.~\eqref{eq:cftotal} represents the effects of unresolved small-scale sea-surface features. Drag due to unresolved surface features, such as capillary waves that cannot be represented on the discretized surface grid, is included through the unresolved roughness length $z_0^u$. The $z_0^u$-dependent contribution in Eq.~\eqref{eq:cftotal} vanishes as $z_0^u\to 0$, corresponding to surfaces that are smooth below the resolved elevation field.

\subsection{\label{sec:delta_selec}Selection of Reference Height $\Delta$ for the pressure drag}

In previous work by \cite{ayala2024}, the reference height $\Delta$ used to define the unperturbed velocity $u_\Delta$ in the form-drag model was taken to be equal to the horizontal LES grid spacing, $\Delta x$. We argue that this choice makes the model overly sensitive to the horizontal grid resolution, since different grids may yield different estimates of the form drag. To reduce this sensitivity, we propose selecting a reference height that depends on the wave surface geometry rather than on the numerical resolution.

In the rough-surface literature, it is common to take the roughness-sublayer height, that is, the height over which the flow is directly influenced by the surface topography, to be approximately 3--5 times the representative height of the roughness elements \citep{chung_review_2021}. In the wind--wave literature, it has similarly been argued that wave effects, or the wave boundary layer, decay above a height of order $1/k$ \citep{sullivan_annual2010,hao_specwave_2019}, where $k$ is the wavenumber. Moreover, analytical studies \citep{ayet2022,Belcher_Hunt_1993} suggest that the inner region in which wave-induced modifications to the turbulent stress are important scales as $z\sim 0.1/k = 0.1a/(ak)$, where $a$ is the wave amplitude and $ak$ is the wave steepness. For wave steepness values in the range $0.1 \le ak \le 0.3$, this estimate gives an inner-region height ranging from $z\sim a$ to $z\sim 3a$. Motivated by these ideas, we choose the reference height as $\Delta = 3H_p$ based on the characteristic dominant positive surface height $H_p$ defined as $H_p = \left\langle \max(0,\eta')^8 \right\rangle^{1/8}$, with $\eta' = \eta - \langle \eta \rangle$ denoting the surface elevation relative to the mean surface level. This method yields a robust estimate of the largest positive excursion of the surface height and is better defined in practice than selecting the single largest height from a finite sample of the surface \citep{meneveau2024}, The pressure-stress model requires the reference velocity $u_\Delta$ to be evaluated at a height where the airflow is only weakly affected by the underlying waves. We therefore argue that using $\Delta = 3H_p$ places the reference point sufficiently far above the  wave surface, while still retaining sensitivity to the characteristic scale of the surface. Smaller-scale waves, which are not adequately represented by $H_p$, are instead accounted for through the unresolved roughness length $z_0^u$.

Finally, because $\Delta$ lies within a region of essentially unperturbed airflow, one generally expects $u_\Delta > c$. As a result, the present model does not explicitly represent critical-layer effects, since the critical layer typically lies well below $\Delta$.
In Appendix \ref{gridmosd_app}, we examine the grid sensitivity of the MOSD model and demonstrate how the selection of the reference height $\Delta$ influences the predicted form drag and the overall robustness of the model.

\subsection{\label{sec:zoU}Estimating Unresolved Surface-Wave Effects Using $z_0^u$}
The MOSD model can account for the drag generated by capillary waves or by wave components that are filtered out at typical LES grid resolutions using $z_0^u$. Although dynamical approaches exist for estimating subgrid surface roughness \citep{ANDERSON_MENEVEAU_2011} and subgrid wave surfaces \citep{Yang_Meneveau_Shen_2013,Aiyer_2024}, here we propose two simpler and less computationally expensive methodologies to estimate \textit{a priori} the equivalent aerodynamic roughness length of unresolved wavy surfaces: 
\begin{enumerate}
    \item Following the work of \cite{Geva}, who proposed a fully rough velocity profile for turbulent airflow over wind-driven waves, the unresolved roughness length can be expressed in terms of the root-mean-square unresolved surface height fluctuation $\eta^\prime_{\mathrm{sgs}}$ as
    \begin{equation}
    z_0^u = \eta^\prime_{\mathrm{sgs}}\,e^{-8.5\kappa}.
    \label{eq:z0_sgs}
    \end{equation}
    It should be noted that this formulation has a limited range of validity, since it was obtained from a fit to experimental data. In those experiments, the authors encountered wave periods in the range $T=0.1$ to $1.5$ sec, corresponding approximately to wavelengths from $\lambda=1.6$ cm to $3.5$ m. Therefore, caution should be exercised when applying this methodology to unresolved surface wavelengths outside that range. In addition, the authors state that their wall-similarity, or fully rough, scaling is appropriate for young wind waves, and they explicitly relate the roughness scaling to cases in which the water surface is effectively fully rough and the drag is dominated by form drag rather than viscous drag.

    \item A second approach is to use the recently developed Surface Wave Aerodynamic Roughness Length (SWARL) model \citep{ayalaSWARL2026}, which estimates the equivalent aerodynamic roughness length induced by ocean waves. This framework combines instantaneous wave-surface elevation maps with an airflow-dependent Reynolds number to estimate the roughness length. Since the SWARL model uses instantaneous surface-height maps, it can used to estimate the roughness associated with the unresolved surface waves whenever the filtered-out surface waves, or the filtered-out surface spectrum, are known.  The model estimates the roughness length as
    \begin{equation}
    z_0^u  = \Delta \exp{(-\kappa\,\Lambda^{-1/2})},
    \label{eq:z0_swarl}
    \end{equation}
    where $\Delta=3H_p$, $\kappa=0.4$ is the von Kármán constant, and $\Lambda$ is a wave-geometry- and flow-dependent factor. Additional details of the model are provided in Appendix \ref{swarl_app}.
\end{enumerate}

Although MOSD and SWARL share some of the same underlying physical considerations, SWARL is intended to provide an equivalent roughness length and therefore represents the influence of waves in a phase-averaged sense, whereas MOSD retains a more explicit dependence on the instantaneous wave geometry. In the present work, our goal is to develop a tractable phase-resolved wall-modeled LES (WMLES) framework for flow over waves.

In this work, we use both representations, in  the present study. Equation~\ref{eq:z0_sgs} is most useful when only statistical information about the unresolved surface is available, such as $\eta^\prime_{\mathrm{sgs}}$. By contrast, when the full unresolved surface field, or its spectrum, is available, the SWARL model provides a more detailed estimate of the corresponding aerodynamic roughness.  We therefore use the SWARL model whenever the filtered-out surface is known, and Eq.~\ref{eq:z0_sgs} when only statistical information about the unresolved wave field is available.

\section{\label{sec:stokes_waveind} Applications in LES: Stokes Waves and Wave-Induced Motions}
In this section, we use the MOSD model to examine two wave-related effects: the possible differences between Stokes and sinusiodal waves, and wave-induced motions in the flow. We compare the results from WMLES using the MOSD model to laboratory experimental data. In all our large eddy simulations, we use a computational domain of size $L_x \times L_y \times L_z = 6\lambda \times 3\lambda \times \delta$, where $\lambda$ is the wavelength. Because the wavelength in all cases is of the same order as the boundary-layer height, and in some cases up to twice as large, this choice of domain size is expected to minimize artificial finite-domain effects on the turbulent flow statistics \citep{Sathe_Giometto_2024}. All simulations are performed on a grid of size $N_x \times N_y \times N_z = 128 \times 64 \times 24$. At this resolution, each wavelength is represented by at least 20 grid points, the grid aspect ratio satisfies $\Delta x/\Delta z \approx 3$, and the maximum surface elevation satisfies $\max[\eta] = a \leq 0.99\Delta_{\rm m}$, where the LES matching height is defined as $\Delta_{\rm m} = 2.5\Delta z$, corresponding to the height of the third vertical grid point in the staggered grid in which the first grid-points of the horizontal velocities are located at a height $\Delta z/2$.  The top boundary of the numerical domain is treated as stress-free, the horizontal boundaries are periodic, and the bottom boundary is modeled using the MOSD wall model. Additional details regarding the LES code (LESGO) are provided in Appendix \ref{lesgo_app}.

\subsection{\label{sec:stokes}Stokes Waves}
In this section we apply MOSD to account for the drag generated by a weakly nonlinear wave, namely a Stokes wave. Similar to a sinusoidal wave, a Stokes wave retains a monochromatic signature (a single dominant wave mode) but it exhibits steeper slopes and sharper crests and troughs. Based on the usual scaling for wave-induced drag \citep{wu_popinet_deike_2022}, $F_D \sim (ak)^2$, one expects a Stokes wave to produce larger drag than a sinusoidal wave with the same amplitude and wavelength. In our previous work \cite{ayala2024}, we compared MOSD-based LES of a third-order Stokes wave against a wave-resolved DNS \citep{wu_popinet_deike_2022} and found that the model underpredicted the mean drag force. We argued that this discrepancy may have arisen because the comparison velocity profiles were spatially averaged at specific instants rather than both spatially and temporally averaged. In the present study, we instead compare our WMLES with recent experiments of a developing boundary layer over paddle-generated waves \citep{zein_waveJFM_2026}, where the authors report that the paddle-generated waves are better represented by a Stokes-like profile than by a purely sinusoidal wave. The experiments considered four paddle-generated wavelengths ranging between $\lambda = 0.124$--$0.41$ m and, for each wave case, three freestream velocities, $u_\infty \approx 2, 4,$ and $6$ m/s were considered. Table \ref{tab:table1} summarizes
the relevant wave and wind parameters from the experiments.

In our LES, the free-surface elevation $\eta$ is prescribed as a third-order Stokes wave:
\begin{equation}
    \eta(x,y,t)=a\cos\theta+\frac{1}{2}(ak)a\cos(2\theta)+\frac{3}{8}(ak)^2a\cos(3\theta),
\end{equation}
where $\theta = kx - \omega t$. 

\begin{table}
\caption{\label{tab:table1} Wave and wind flow parameters for Stokes waves  }
\begin{ruledtabular}
\begin{tabular}{ccccccc}
Case & $u_\infty$ (m/s) & $a$ (m) & $\lambda$ (m) & $u_*$ (m/s) & $\delta$ (m) & $z_0^u$ (m) \\
\hline
S1 & 2.41 & 0.118 & 0.410 & 0.124 & 0.148 & $1.185 \times 10^{-5}$\\
S2 & 3.87 &  &  & 0.211 & 0.148 & $2.035 \times 10^{-5}$\\
S3 & 5.42 &  &  & 0.308 & 0.159 & $1.771 \times 10^{-5}$\\

S4 & 2.41 & 0.150 & 0.313 & 0.139 & 0.146 & $1.249 \times 10^{-5}$\\
S5 & 3.87 &  &  & 0.230 & 0.134 & $1.357 \times 10^{-5}$\\
S6 & 5.42 &  &  & 0.353  & 0.142 & $2.203 \times 10^{-5}$\\

S7 & 2.41 & 0.092 & 0.189 & 0.164 & 0.130  & $1.435\times 10^{-5}$\\
S8 & 3.87 &  &  & 0.285 & 0.131 & $1.274 \times 10^{-5}$\\
S9 & 5.42 &  &  & 0.384 & 0.139 & $1.693 \times 10^{-5}$\\

\end{tabular}
\end{ruledtabular}
\end{table}

The authors also observed that the waves were not perfectly smooth, but were instead covered by small wind-driven ripples superposed on the primary paddle-generated wave. Due to experimental resolution limitations, only the root mean square (r.m.s.) amplitude of the ripples could be obtained for each case. To estimate the contribution of these unresolved surface fluctuations, we first define the fluctuation field as $\eta' = \eta - \eta_{\rm st}$, where $\eta$ is the instantaneous experimentally measured surface elevation and $\eta_{\rm st}$ is the corresponding fitted Stokes-wave profile. The fluctuation field is then high-pass filtered using a cutoff scale $\Delta = 2\Delta x$. The r.m.s. subgrid surface height is subsequently computed as $\eta'_{{\rm sgs}} = \sqrt{\langle \eta'^2 \rangle}$, from which the equivalent unresolved surface roughness length $z_0^u$ is estimated using Eq.~\ref{eq:z0_sgs}. The resulting values of the subgrid surface roughness for each case are also reported in Table~\ref{tab:table1}.  We note that the experiments correspond to a zero-pressure-gradient developing boundary layer, in which the freestream velocity $u_\infty$ drives the airflow above the waves. In contrast, in our LES the flow is driven by a constant pressure gradient, $-dP_\infty/dx = u_*^2/\delta$, where the domain height $\delta$ is set equal to the experimental boundary-layer thickness $\delta_{99}$ and the friction velocity $u_*$ is prescribed from the experimental data.
 
\begin{figure*}
\includegraphics[width=\textwidth]{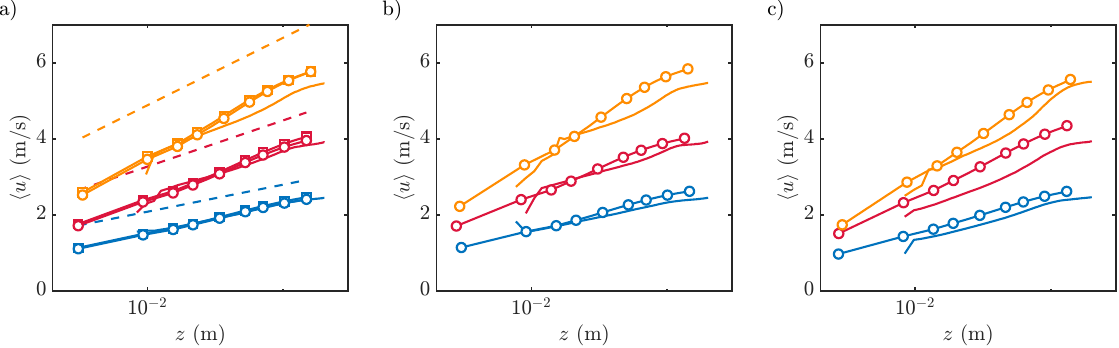}
\caption{\label{fig:fig1} Mean streamwise velocity profiles for the a) $\lambda=0.41$m cases (S1,S2,S3), b) $\lambda=0.31$m cases (S4,S5,S6) and c) $\lambda=0.19$m cases (S7,S8,S9). Each color represents a different freestream velocity: $u_\infty=2.41$m/s (Blue), $u_\infty=3.87$m/s (Red) and $u_\infty=5.42$m/s (Yellow). Solid lines: Experiments from \cite{zein_waveJFM_2026}. Circles: WMLES Stokes wave. Square: WMLES Sinusoidal wave. Dashed line: Log-law profile $\kappa^{-1}\log{(z/z_0^u)}$.}
\end{figure*}

Figure \ref{fig:fig1} compares the mean streamwise velocity profiles obtained from our WMLES with the MOSD model with those measured in the experiments. Overall, the WMLES results for the Stokes-wave cases show good agreement with the experimental data across the range of conditions considered. One notable exception occurs for the shortest-wavelength cases, $\lambda = 0.19$ m (cases S7--S9; Fig.~\ref{fig:fig1}c), particularly case S8, for which the discrepancy is more pronounced. Given the generally good agreement observed for the remaining cases, the mismatch for these cases may be associated, in part, with uncertainties in representing the small-scale ripple contribution. Although the aerodynamic effect of the unresolved ripples is included through the roughness length $z_0^u$, this quantity is estimated from the r.m.s. amplitude of the high-pass-filtered surface fluctuations using Eq.~\ref{eq:z0_sgs}. Thus, the unresolved ripple field is represented through a single equivalent roughness length rather than its full surface geometry. Experimental resolution limitations may lead to incomplete characterization of the smallest-scale surface fluctuations, while the roughness parameterization itself does not retain information regarding their wavelength, slope, or spatial distribution. These effects may lead to an inaccurate estimate of the additional drag associated with the ripple field and could contribute to the discrepancies observed for the $\lambda=0.19$ m cases.

In Fig.~\ref{fig:fig1}a, we also include the logarithmic profiles (dashed lines) based only on the unresolved surface roughness, showing that the MOSD model contributes additional drag through the resolved form-drag component. For the same case, $\lambda = 0.41$ m, we additionally simulated the corresponding wind--wave condition using a sinusoidal wave; those results are shown in the figure as square symbols. Although the differences are small, they are nevertheless clear: the Stokes-wave case exhibits a slightly larger vertical offset in the mean velocity profile, indicating the expected somewhat larger drag than the corresponding sinusoidal-wave case. 

\begin{figure*}
\includegraphics[width=\textwidth]{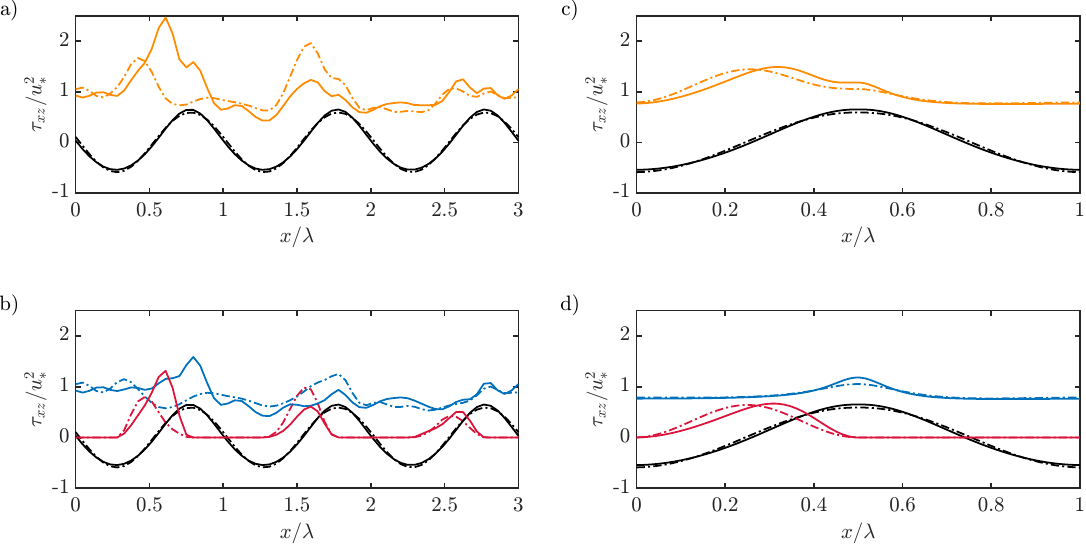}
\caption{\label{fig:fig2} a) and b) Instantaneous profiles of the wall stress components for S1 case. c) and d) Phase-averaged profiles of the wall stress components for S1 case. For all figures, the wave drag component $\tau_{xz}^{w-p}$ is shown as red line, the $\tau_{xz}^{w-u}$ component is shown as blue line, the total stress is shown as yellow line. Solid line: Stokes wave. Dashed-dot line: Sinusoidal wave. The wave profile are shown as black line and are not to scale.  }
\end{figure*}

To highlight the differences between the Stokes and sinusoidal waves more clearly, we plot the instantaneous profiles of the wall-stress components of the MOSD model  in Fig.~\ref{fig:fig2}a,b. The wave-drag component is larger for the Stokes wave than for the sinusoidal wave over the windward portion of the wave. For both cases, the wave-drag component vanishes over the leeward side, consistent with the assumption of flow separation or streamline displacement in that region imposed by the MOSD model. To facilitate this comparison, we use a triple decomposition \citep{hussain_reynolds_1970}:
\begin{equation}
    f(x,y,z,t) = \langle f \rangle(z) + \tilde{f}(x,z,\phi) + f'(x,y,z,t)
\end{equation}
where $\langle f \rangle$ is the ensemble-averaged component, $\tilde{f}$ is the wave-induced component that depends on phase $\phi$, and $f'$ is the uncorrelated turbulent fluctuation. The mean component $\langle f \rangle$ is obtained by averaging over both horizontal directions and time. The wave-induced component is defined as $\tilde{f} = f_\phi - \langle f \rangle$, where $f_\phi$ is estimated as
\begin{equation}
    f_\phi(x,z,\phi) = \frac{1}{N}\sum_{n=0}^{N} \langle f(x,y,z,(\phi/2\pi+n)T) \rangle_y
    \label{eq:phase_avg}
\end{equation}
where $T$ is the wave period and the subscript $y$ denotes averaging in the spanwise direction.

The phase-averaged wall-stress components are shown in Fig.~\ref{fig:fig2}c,d. Here, the phase-averaged stress is first computed using Eq.~\ref{eq:phase_avg} and is then additionally averaged over all wavelengths. We observe that the peak of the wave-drag component for the Stokes wave is slightly shifted relative to that of the sinusoidal wave. This shift arises because the drag model depends on the local surface gradient, and for a Stokes wave the location of the maximum slope is itself slightly shifted relative to that of a sinusoidal wave. The viscous component, $\tau_{xz}^{w-u}$, also exhibits a larger peak for the Stokes wave. This follows from the dependence of the viscous stress on the local surface elevation through the term $\log[(\Delta_{\rm m}-\eta)/z_0^u]$. Because the crest of the Stokes wave is higher than that of the sinusoidal wave, the quantity $(\Delta_{\rm m}-\eta)$ is smaller near the crest. For fixed $\Delta_{\rm m}$ and $z_0^u$, this reduces the logarithmic denominator and therefore increases the stress locally.

\subsection{\label{sec:waveind}Wave-Induced Motions}

\begin{figure*}
\includegraphics[width=15cm]{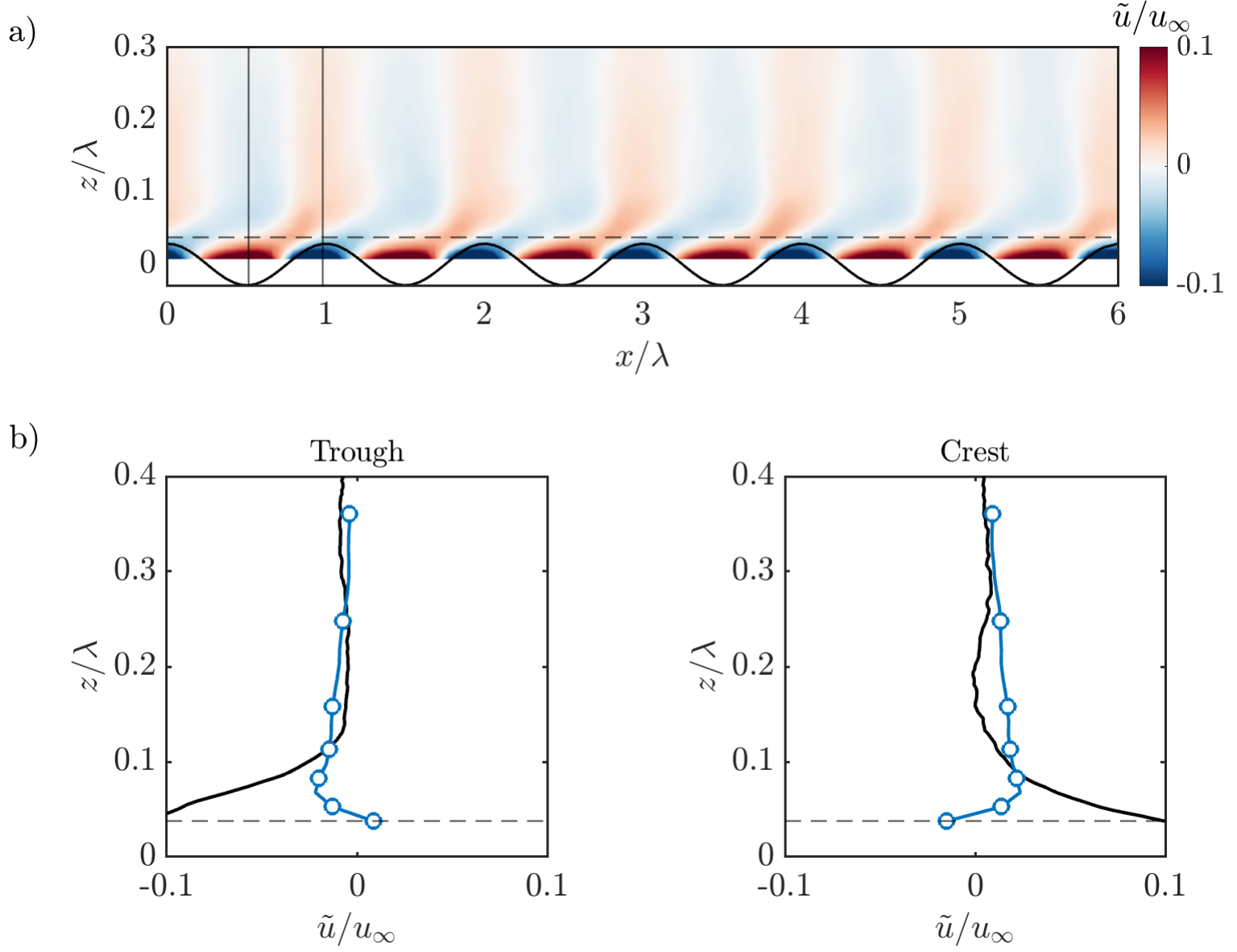}
\caption{\label{fig:fig3} a) Normalized streamwise wave-induced velocity for the S1 case. Vertical black lines represent the location of the crest and through. b) Vertical profiles of the streamwise wave-induced velocity at the through and at the crest for S1 case. Black line: Experiments by \cite{zein_waveJFM_2026}. Blue line: WMLES. For all figures, the horizontal dashed line represents the location of the matching location height.}
\end{figure*}

In our previous work \citep{ayala2024}, we demonstrated the phase-resolving capability of the MOSD model by applying phase averaging to the velocity field and analyzing the resulting wave-induced motions. Here, we continue that assessment by comparing our WMLES results with the experimental data. Figure~\ref{fig:fig3}a shows contours of the normalized streamwise wave-induced velocity for the wave phase corresponding to the surface position shown in the figure. The dashed line denotes the matching height $\Delta_{\rm m}$; therefore, data below this line may be considered to lie "inside" the wave. We emphasize that the flow field below the dashed line is strongly influenced by the imposed boundary condition, that is, the MOSD model, whereas the flow above this height results solely from the filtered momentum equations.

Above the matching location, the streamwise wave-induced velocity is positive near the wave crest and negative near the wave trough, consistent with the findings of \citep{sullivan_2000,yousefi_2020}. Below the matching location, however, we observe an earlier onset of flow acceleration in the trough region and flow deceleration near the crest. This near-surface behavior is expected because the MOSD model imposes enhanced drag near the wave crest (see Fig.~\ref{fig:fig2}d), which locally decelerates the flow.

To further examine this behavior, we consider the vertical profiles of the streamwise wave-induced velocity at the crest and trough, shown in Fig.~\ref{fig:fig3}b. Near the matching height, the amplitude of the wave-induced motion is underpredicted by approximately a factor of 3--4 relative to the experiments. Similar discrepancies are observed for the vertical wave-induced velocity, suggesting that the model underpredicts the overall strength of the near-surface wave-coherent motions. This behavior is consistent with the MOSD formulation, which imposes a boundary condition only on the horizontal velocity components $(u,v)$, while the vertical velocity ($w$) is generated indirectly through mass conservation. Importantly, the agreement improves with increasing distance from the surface, and the model captures the qualitative structure of the wave-induced flow above the matching region. Future improvements should therefore focus on better representing the near-surface wave-induced motions.

\begin{figure*}
\includegraphics[width=15cm]{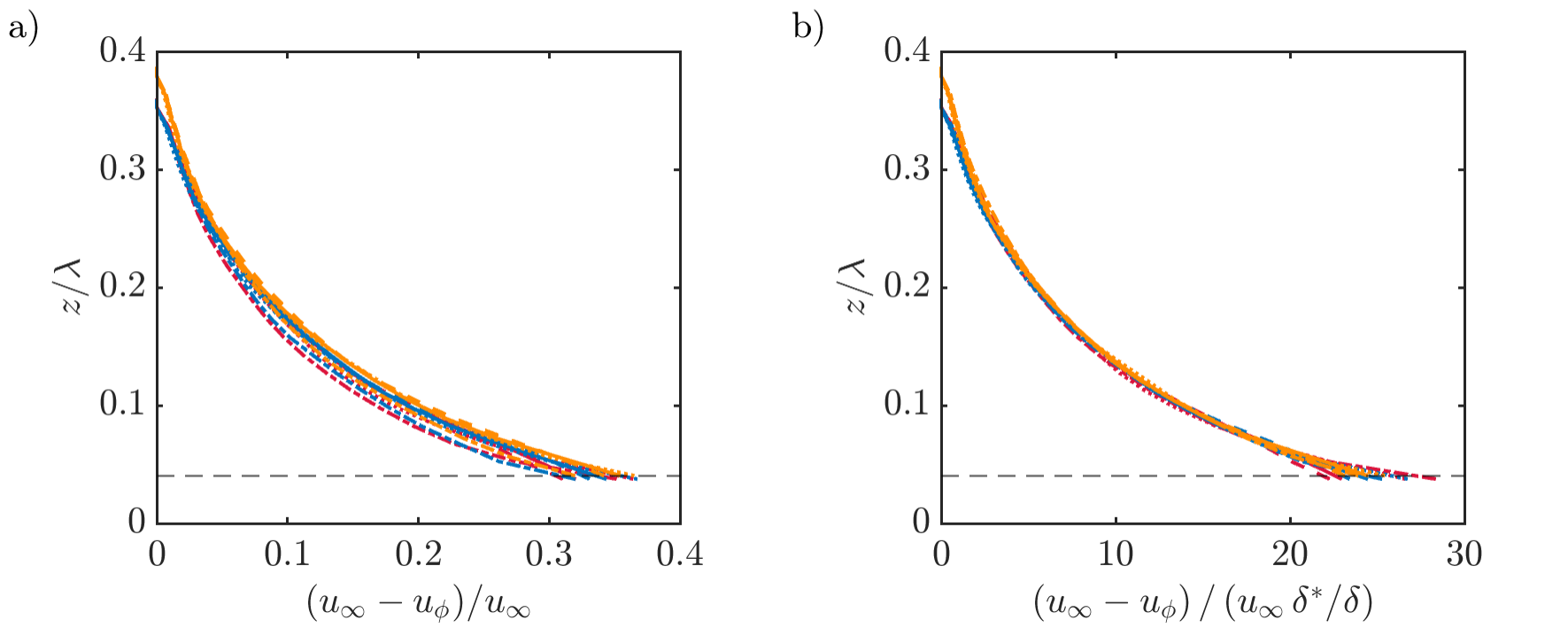}
\caption{\label{fig:fig4} a) Vertical profiles of the phase-averaged velocity deficit at several phases for S1,S2 and S3 cases. b) Scaled vertical profiles of the phase-averaged velocity deficit at several phases for S1,S2 and S3 cases. For all figures, the horizontal dashed line represents the location of the matching location height.}
\end{figure*}

The work of \cite{zein_waveJFM_2026} showed that the phase-averaged streamwise velocity at different wave phases can be collapsed by scaling the phase-averaged velocity deficit with the ratio $\delta^*/\delta$, following the approach originally proposed by \cite{zagarola1998mean}. The displacement thickness $\delta^*$ is defined as
\begin{equation}
    \delta^*=\int_0^\delta \left(1-\frac{u_\phi}{u_\infty}\right)\,dz.
\end{equation}
Figure~\ref{fig:fig4}a shows the phase-averaged velocity deficit at several wave phases, $x/\lambda=0.25, 0.5, 1,$ and $1.25$, corresponding to the ascending face, trough, crest, and descending face, respectively. Far from the surface, the profiles approach zero. Closer to the surface, but still above the matching height, the deficit profiles diverge, in agreement with experimental observations \citep{zein_waveJFM_2026}. Figure~\ref{fig:fig4}b the scaled phase-averaged velocity deficit. Consistent with the experimental results, the scaled profiles collapse well.

\section{\label{sec:spectrum}Broadband Spectrum Waves}

In previous work \citep{ayala2024}, we demonstrated the applicability of the MOSD model to non-monochromatic waves, specifically wave fields composed of two and three distinct modes, as a precursor to more realistic broadband wave spectra. In this section, we extend that analysis by using the MOSD model to simulate turbulent flow over broadband spectrum waves and compare results within the wave-resolved simulations of \cite{Yang_Meneveau_Shen_2013}. We consider three distinct wind--wave scenarios designed to mimic the cases studied by \cite{Yang_Meneveau_Shen_2013}.

For broadband wave fields, the surface elevation $\eta(x,y,t)$ is generated using the standard approach of superposing random-phase traveling waves with a prescribed surface spectrum $S(k_x,k_y)$ \citep{Deskos2021,sullivan2014}:
\begin{equation}
\eta(x,y,t) = \int_{\bf{k}} \sqrt{2S(k_x,k_y)}\, e^{i({\bf{k}\cdot{\bf x}}-\omega t+\phi_{\bf{k}})}\, d^2{\bf{k}},
\label{eq:wave_amp_sepctra}
\end{equation}
where $\phi_{\bf{k}}$ are prescribed random phases, which are kept fixed for all cases. The multiscale surface is generated using the same spectrum adopted by \cite{Yang_Meneveau_Shen_2013}, namely the JONSWAP spectrum \citep{hasselman1973}:
\begin{equation}
S(k_x,k_y) = \frac{\alpha_p}{2k^4}\exp{\left[-\frac{5}{4}\left(\frac{k_p}{k}\right)^2\right]}\, 3.3^\gamma\, D(\theta),
\label{eq:jonswap}
\end{equation}
where $\alpha_p$ is the Phillips constant, $\gamma={\exp{[-{1}/{(2 \varepsilon^2)} ( \sqrt{{k}/{k_p}}-1)^2 }}]$, and the directional distribution is prescribed using the widely used spreading function \citep{hasselmann1980,cartwright1963},
$$
D(\theta)=\frac{2}{\pi}\cos^2(\theta-\theta_w),
$$
for $|\theta|\leq \pi/2$, where $\theta=\arctan(k_y/k_x)$ and $\theta_w$ is the direction of the peak wave. For all cases considered here, we set $\theta_w=0$.

Spatial and temporal gradients of all wave fields are computed using first-order finite differences. For all cases, we use a numerical domain of size
$
L_x \times L_y \times L_z = 12\lambda \times 12\lambda \times 2\lambda.
$
To assess the robustness of the wall model, we simulate each case at two grid resolutions: a coarse resolution of
$
N_x \times N_y \times N_z = 64 \times 64 \times 32,
$
and a finer resolution of
$
N_x \times N_y \times N_z = 128 \times 128 \times 64.
$
All cases satisfy $\Delta x/\Delta z = 3$ and $\max[\eta]\leq 0.99\Delta_{\rm m}$.

We note that the wave-resolved simulation of \cite{Yang_Meneveau_Shen_2013} uses grid spacings of $\Delta x/\lambda_p=0.0416$ and $\Delta y/\lambda_p=0.0156$, corresponding to a resolution, or filter, wavenumber of $k_\Delta=[(\pi/\Delta x)^2 + (\pi/\Delta y)^2]^{1/2}\approx 215/\lambda_p$, whereas our finest WMLES resolution corresponds to $k_\Delta \approx 47/\lambda_p$. Thus, in our WMLES the wave field is necessarily filtered relative to the fully wave-resolved simulation. To account for the drag associated with unresolved subfilter-scale surface features, we include the effects of the unresolved roughness length $z_0^u$, estimated using Eq.~\ref{eq:z0_swarl}. Specifically, we estimate the subgrid surface roughness by reconstructing the portion of the JONSWAP wave spectrum lying between $k_\Delta \approx 47/\lambda_p$ and $k_\Delta \approx 215/\lambda_p$, that is, the range of wave components resolved in the wave-resolved LES but filtered out in our WMLES. In this sense, the procedure is equivalent to applying a high-pass filter to the wave-resolved spectrum in order to isolate the unresolved surface contribution. 

As before, for all simulations, the top boundary is stress-free, the horizontal boundaries are periodic, and the bottom boundary is modeled using the MOSD wall model. The flow is driven by a constant pressure gradient,
$-dP_\infty/dx = u_*^2/L_z$. The wave and flow parameters for each wind--wave scenario are summarized in Table~\ref{tab:table2}. 

\begin{table}
\caption{\label{tab:table2} Wave and wind flow parameters for flow over broadband spectrum waves. $H_sk_p$ is the wave steepness, where $H_s=4\sqrt{S(k_x,k_y)}$ is the significant wave height. $k_p$ and $c_p$ are the peak wavenumber and phase velocity, respectively. $Re_\tau=u_*\lambda_p/\nu$ is the turbulent Reynolds number.  }
\begin{ruledtabular}
\begin{tabular}{ccccccc}
Case & $\alpha_p$ & $H_sk_p$ & $k_p$ (m) & $c_p/u_*$ & $Re_\tau$ &$z_0^u/\lambda_p$ \\
\hline
B1c & 0.0267 & 0.36 & 1.38 & 6 & 4096 &$9.43 \times 10^{-5}$\\
B1f & & & & & & $5.72 \times 10^{-5}$\\

B2c & 0.0191 & 0.28 & 0.50 & 10 & 4096 &$6.31 \times 10^{-5}$\\
B2f & & & & & & $5.25 \times 10^{-5}$\\

B3c & 0.0129 & 0.20 & 0.15 & 18 & 4096 &$6.19\times 10^{-5}$\\
B3f & & & & & & $5.35\times 10^{-5}$\\

\end{tabular}
\end{ruledtabular}
\end{table}

\begin{figure*}
\includegraphics[width=\textwidth]{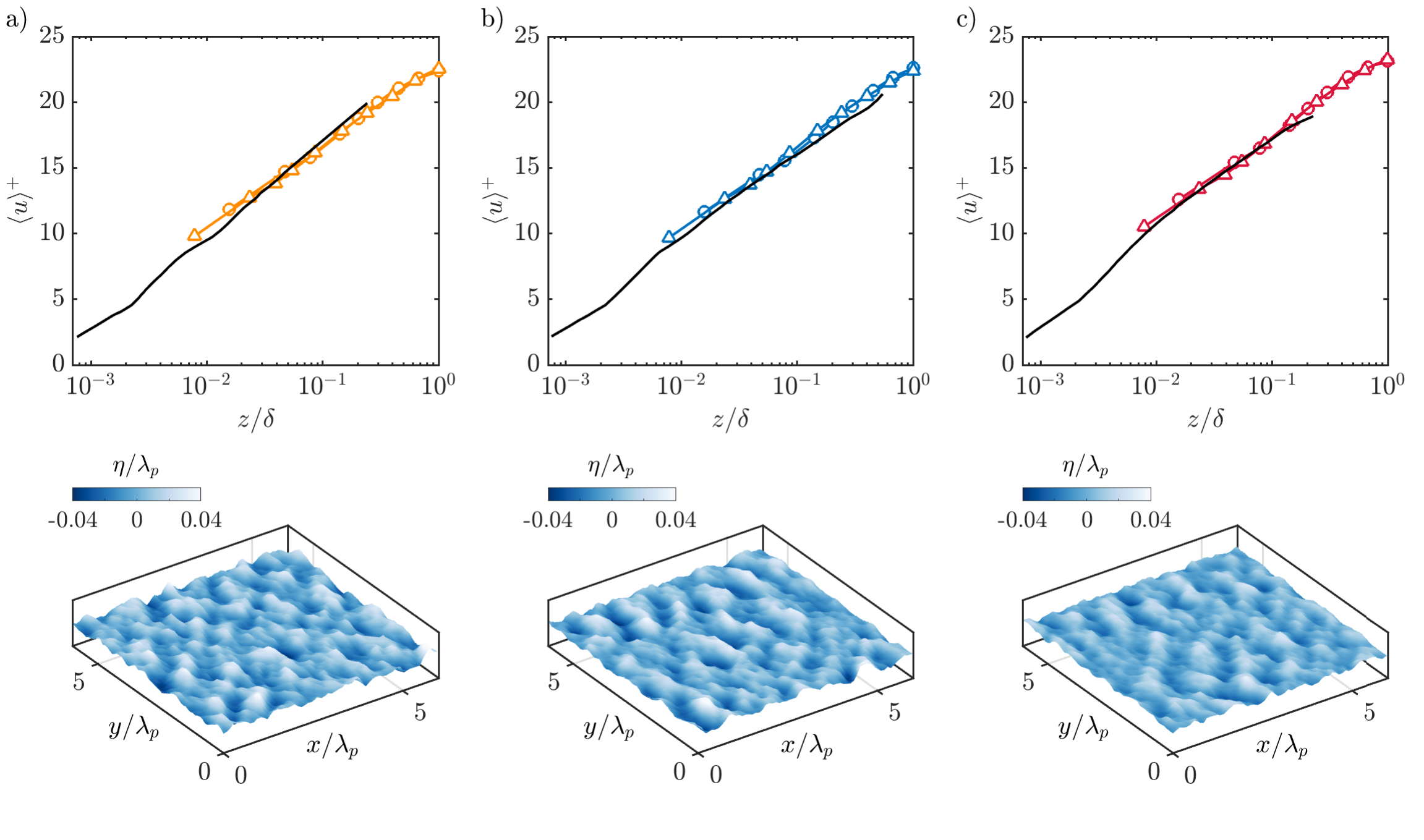}
\caption{\label{fig:fig5} Mean streamwise velocity profiles for a) case B1, b) case B2 and c) case B3. The triangle symbol represents the coarse resolution while the circle symbol represents the fine resolution. Black line: Wave-resolving LES from \citep{Yang_Meneveau_Shen_2013}. A snapshot of the corresponding broadband wavefield is shown below each velocity profile.}
\end{figure*}

Figure~\ref{fig:fig5} shows the mean streamwise velocity profiles for all broadband-wave cases, comparing our WMLES results with the wave-resolved LES of \cite{Yang_Meneveau_Shen_2013}. For all cases, the WMLES using MOSD shows good agreement with the wave-resolved LES data, regardless of the grid resolution used. These results validate the accuracy and   applicability of the model for predicting momentum transfer over broadband waves and its robustness across different levels of wave-field resolution.

\section{\label{sec:owbl_broadband}Offshore Wind Farm Boundary Layer over Broadband Spectrum Waves}

We further assess the applicability of the MOSD model by using it in WMLES  of offshore wind farms. We compare the results with the wave-resolving LES of \citep{yang2014large}, which examined the influence of realistic broadband spectrum waves on the flow statistics and coherent structures of an offshore wind-farm boundary layer. The reference study considered three distinct wind-farm array configurations under fully developed periodic conditions, all forced by the same prescribed broadband wave field. Here, we reproduce those cases as closely as possible in order to enable a direct comparison with the wave-resolving simulations. To this end, we adopt the same physical parameters and numerical setup as in \cite{yang2014large}. In this way, the comparison serves not only as a validation of the model in a more complex setting, but also as a demonstration of its tractability and robustness for large-scale offshore wind-energy applications.

For the broadband wave field, the surface elevation is prescribed as a superposition of random-phase traveling waves with a JONSWAP spectrum, following Eqs.~\ref{eq:wave_amp_sepctra}--\ref{eq:jonswap}. The Phillips constant is set to $\alpha = 0.012$, the peak wavenumber is $k_p = 0.1~\mathrm{m}^{-1}$, and the peak-wave direction is taken as $\theta_w = 0$. The wind turbines are represented using the standard actuator disk model (ADM) \citep{calaf2010large,shapiro2019}. Each turbine has hub height $\delta_h = 100$ m and rotor diameter $D = 100$ m. The wind-farm arrays considered are $3$ columns $\times N_{\rm row}$rows , with $N_{\rm row}=2,3,4$ The corresponding streamwise turbine spacings are $s_x = (L_x/N_{\rm row})/D = 10.5, 7.0,$ and $5.25$ for the three cases, respectively. The spanwise spacing is fixed at $s_y = (L_y/3)/D = 5.0$. For all cases, we use a numerical domain of size $L_x \times L_y \times L_z = 2100 \times 1500 \times 1000$ m and a grid resolution of $N_x \times N_y \times N_z = 192 \times 128 \times 192$. All cases satisfy $\Delta x/\Delta z = 3$ and $\max[\eta]\leq 0.99\Delta_m$.

\begin{figure*}
\includegraphics[width=\textwidth]{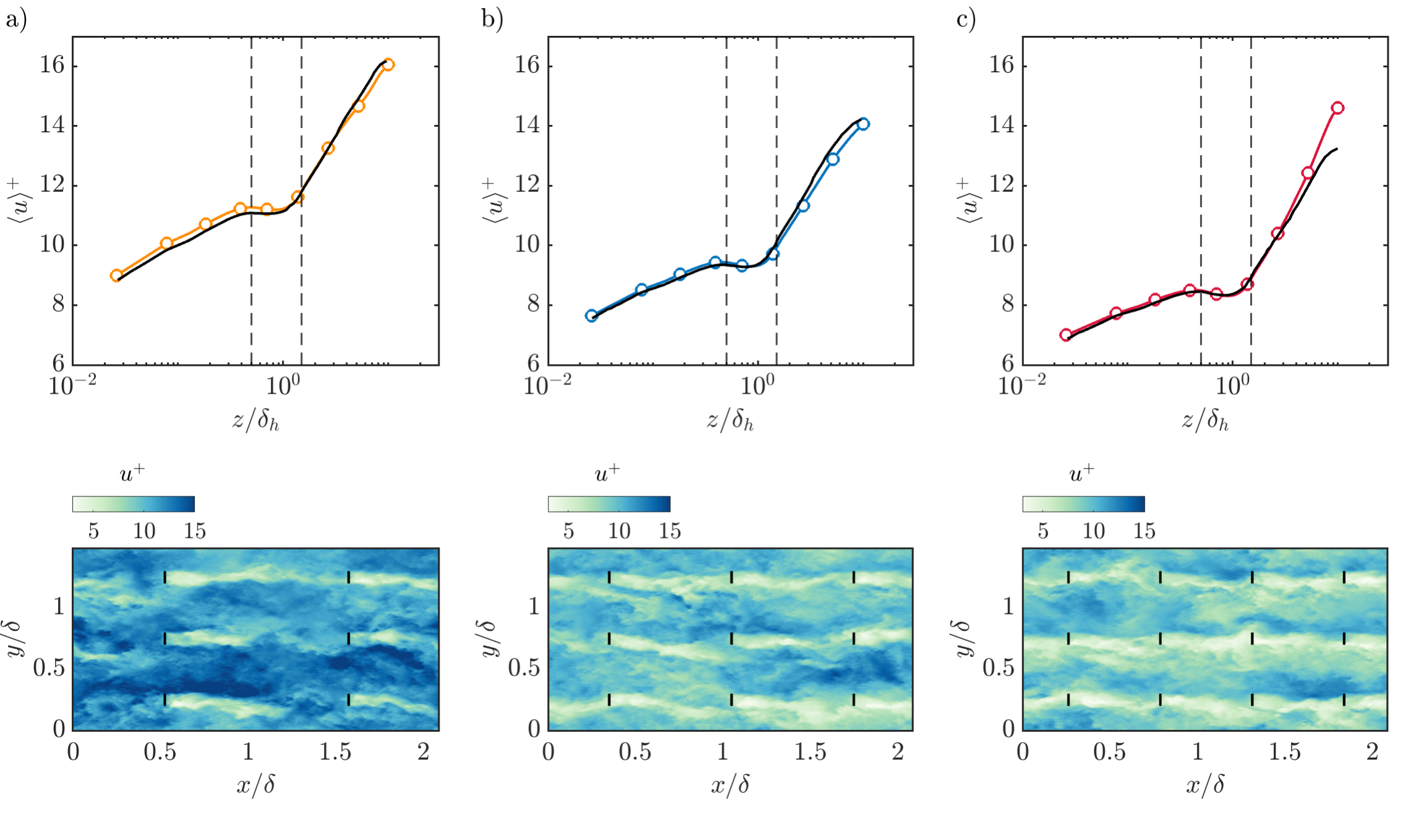}
\caption{\label{fig:fig6} Mean streamwise velocity profiles for a) case BT1, b) case BT2 and c) case BT3. Triangle symbols represents the coarse resolution while   circles  represents the fine resolution. Black line: Wave-resolving LES from \citep{yang2014large}. Below each figure a snapshot of the corresponding flow field at hub height is shown.}
\end{figure*}

Figure~\ref{fig:fig6} shows the mean streamwise velocity profiles for all cases. Beneath each panel, we also include a contour plot of the instantaneous streamwise velocity in a horizontal plane at hub height, illustrating the imprint of the wind turbines on the surrounding flow field. Overall, the MOSD-based LES reproduces the mean velocity profiles from the wave-resolving simulations well across the full depth of the wind-farm boundary layer, both below and above the turbine region. This level of agreement is obtained for all three cases considered in this section. Although the WMLES is performed using the same numerical resolution as the wave-resolving simulations, the present approach obtains these results at a significantly lower computational cost.  The effects of the waves are introduced through the MOSD  model, allowing the important phase-dependent, wave-induced modifications to the flow to be captured without the need to explicitly resolve the wave field with a wave-following or coordinate-transformed grid. 

\section{\label{sec:owbl_realistic}Offshore Wind Farm Boundary Layer under Realistic Inflow Conditions}

\begin{figure}
\includegraphics[width=8cm]{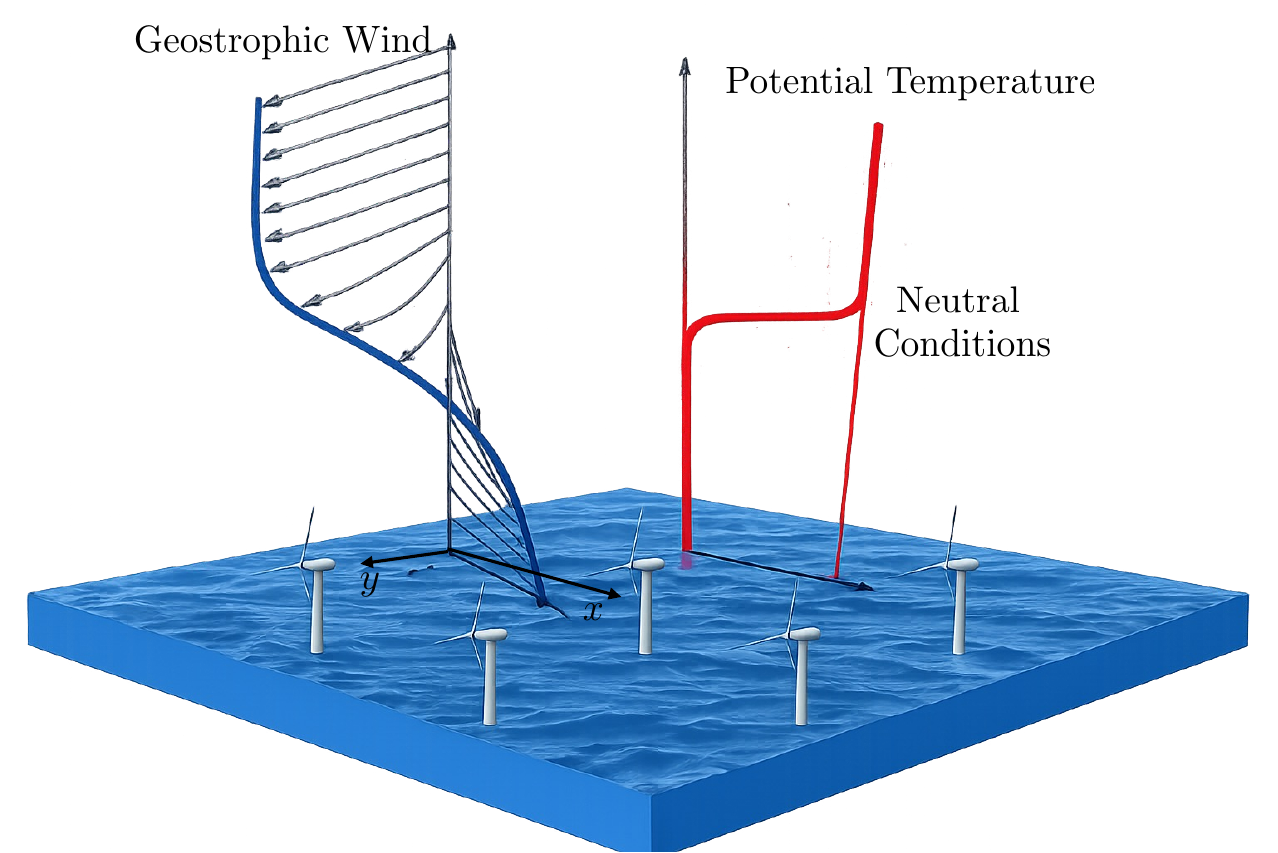}
\caption{\label{fig:fig7} Schematic representation of an offshore wind farm under realistic atmospheric conditions, namely over broadband ocean waves and under a conventionally neutral stability condition. The design of this figure is inspired on a schematic presented \cite{Allaerts_Meyers_2017}.} 
\end{figure}

In this section, we consider an offshore wind-farm boundary layer under conditions that are more representative of the marine atmospheric boundary layer (see Fig.~\ref{fig:fig7}). In particular, instead of using periodic boundary conditions in the streamwise direction, we include a separate inflow simulation to avoid having to assume fully developed conditions.  Our objective is primarily exploratory: we seek to determine whether ocean waves of sufficiently large scale can measurably influence the power production of a fixed-bottom offshore wind farm that includes inflow effects.

In the wind-wave literature, it has been argued that wave effects decay above a height of order $1/k$, or equivalently on the order of the wavelength scale \citep{sullivan_annual2010,hao_specwave_2019}. This suggests that sufficiently long waves may influence the flow not only near the surface, but also at elevations relevant to turbine operation. Recent wave-resolving simulations by \cite{Ciri2024} showed no clear wave-induced signature in the tangential loads of a wind turbine under growing-wave conditions, although that study considered wavelengths only up to approximately $50$ m for a turbine with hub height of $90$ m. By contrast, other wave-resolving simulations \citep{yang_offshorefarm_2014,yang_offshoreturbine_2022} have shown that, under swell conditions, wave-induced signatures can appear in turbine power production.

Motivated by these findings, we perform a study designed to isolate whether large-scale waves can influence wind-farm power production, while remaining within the intended range of applicability of the MOSD model. To this end, we simulate three scenarios for a $2\times4$ fixed-bottom offshore wind farm. The three cases differ only in the wind--wave condition imposed: a no-wave case, a monochromatic sinusoidal-wave case, and a broadband-spectrum-wave case. For the wave cases, we prescribe a peak wave speed of $c_p=12$ m/s, which corresponds, through the deep-water dispersion relation, to a peak wavelength of approximately $\lambda_p\approx 92$ m. This wavelength is intentionally chosen to be of the same order as the turbine hub height, so that the possible impact of wave-induced flow modifications on turbine performance can be examined directly. For the broadband-wave case, the surface elevation is generated from a JONSWAP spectrum with Phillips constant $\alpha_p=0.0056$ and peak-wave direction $\theta_w=0$, yielding a significant wave height of approximately $H_s\approx2.4$ m. In the no-wave scenario (case R1), we set $\tau^{\mathrm{w-p}}_{i3}=0$, so that the lower boundary stress is due only to the prescribed unresolved roughness. The wave conditions for all three scenarios are summarized in Table~\ref{tab:table3}. We note that the same unresolved roughness length is prescribed in all cases so that the comparison isolates the effect of the resolved wave field.

\begin{table}
\caption{\label{tab:table3} Wave and wind flow parameters for offshore wind farm boundary layer under realistic conditions. For the R2 case, the significant wave height is the sinusoidal wave amplitude ($H_s=a$) }
\begin{ruledtabular}
\begin{tabular}{cccccc}
Case & $\alpha_p$ & $H_s$ (m) & $\lambda_p$ (m) & $z_0^u$ (m) \\
\hline
R1 & - & - & -  &$2 \times 10^{-4}$\\
R2 & - & 2.94 & 92  &$2 \times 10^{-4}$\\
R3 & 0.0056 & 2.4 & 92  &$2 \times 10^{-4}$\\
\end{tabular}
\end{ruledtabular}
\end{table}

We drive the flow using a geostrophic wind of magnitude $G=20$ m/s. At each time step, a proportional--integral (PI) controller is used to adjust the geostrophic forcing so that the wind remains aligned with the streamwise direction and exhibits zero wind veer at hub height \citep{sescu2014,ghanesh2022}. We note that the selected geostrophic wind speed is more characteristic of strong extratropical-storm or frontal-passage conditions than of typical offshore operating conditions. Nevertheless, this choice serves two purposes: it maintains the wave field in a growing-wave regime, with $c_p/U_{10}<1.2$ \citep{sullivan2014}, where $U_{10}$ is the velocity at 10 m height, rather than in a swell-dominated regime, and it allows us to consider a large wavelength that remains within the intended range of applicability of the MOSD model. We again emphasize that the MOSD model is not designed to capture swell-dominated conditions reliably.

Conventionally neutral stratification is introduced by prescribing an initial potential-temperature profile with a capping inversion, such that the boundary layer is neutral below the inversion and stably stratified above it. The inversion height is set to 1 km above the surface. Below the inversion, the potential temperature is initialized to $\theta=265$ K, equal to the reference temperature $\theta_0$. Above the inversion, the potential temperature increases with a lapse rate of $0.001$ K/m. To maintain conventionally neutral conditions within the MABL, we impose a zero-buoyancy-flux boundary condition at the lower boundary.

The simulations are carried out using two computational domains within the concurrent-precursor framework of \citep{stevens2014}. The precursor domain generates the turbulent inflow corresponding to conventionally neutral flow over waves, while the turbine domain contains the wind-farm array. Both domains use periodic boundary conditions in the spanwise direction and a stress-free upper boundary. In the wind-farm domain, a fringe region of size $L_x/8$ is used to enforce consistency between outflow and inflow at the periodic boundary. In the precursor domain, a shifted periodic boundary condition is employed, in which the flow field in the fringe region is shifted in the spanwise direction to prevent persistent locking of large-scale turbulent structures \citep{munters2016}.

We consider a 4 row by 2 column wind-farm array in which each turbine has hub height $\delta_h=90$ m and rotor diameter $D=126$ m. The turbines are spaced $s_x=7.0D$ in the streamwise direction and $s_y=5.0D$ in the spanwise direction. All turbines use a local thrust coefficient $C_T'=1.33$ and are represented using the standard ADM \citep{shapiro2019}. For all scenarios considered in this section, both the precursor and turbine domains have size $L_x \times L_y \times L_z = 6048 \times 1260 \times 2000$ m and resolution $N_x \times N_y \times N_z = 656 \times 128 \times 400$. The horizontal dimensions are chosen to accommodate the $2\times4$ turbine array, the prescribed turbine spacing, and the fringe region. Each domain contains approximately 66 peak wavelengths resolved by about 10 streamwise grid points. We note that the vertical extent, $L_z=2$ km, is considerably smaller than that used in studies specifically designed to resolve wind-farm-induced atmospheric gravity waves \citep{Allaerts_Meyers_2017,Lanzilao_Meyers_2024}. The present study does not aim to resolve or quantify these large-scale gravity-wave dynamics; instead, we focus on the lower MABL and the direct influence of ocean waves on the turbine-rotor region. To limit spurious reflections from the upper boundary, a sponge layer is applied between $0.75L_z$ and $L_z$, with an inverse relaxation time scale of $3.9\times10^{-3}\ \mathrm{s}^{-1}$.

\subsection{\label{sec:general_obs}General Observations}

\begin{figure*}
\includegraphics[width=\textwidth]{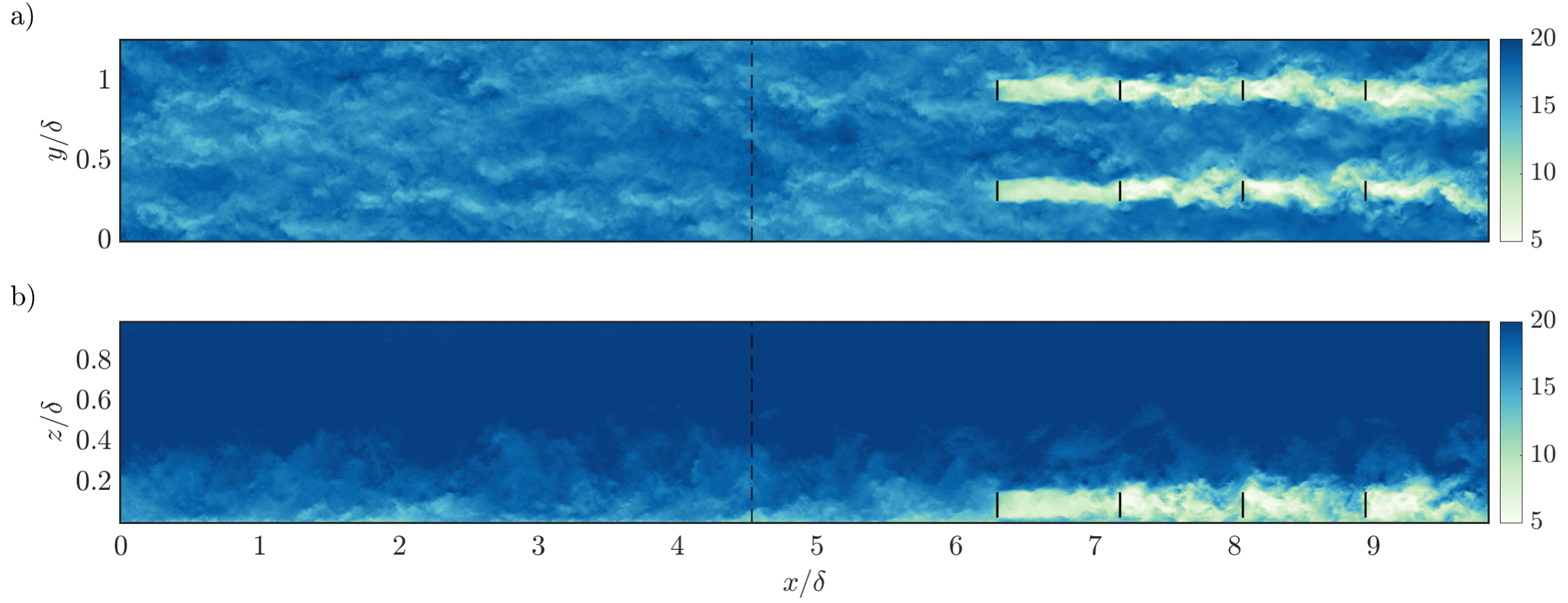}
\caption{\label{fig:fig8} Instantaneous contours of the normalized velocity ($u/u_*$) at a) hub height (top view) and at the b) middle of the domain (side view). The precursor and turbine domain are shown stitched together by removing the last $1/4$ of the precursor domain (the fringe region used for lateral shifting) and the last $1/8$ of the wind farm domain (the fringe region of the wind farm domain). Vertical dashed line represents the merging location between both domains. Black thick lines represent the wind turbines. We define $\delta=L_z=2000$m.}
\end{figure*}

Figure~\ref{fig:fig8} shows instantaneous velocity contours in both the precursor and turbine domains. The turbulent MABL inflow generated in the precursor domain feeds into the wind-farm domain, where each turbine row produces a wake that progressively modifies the downstream flow. As expected, the large-scale wake dynamics are clearly visible in these instantaneous fields. By contrast, the influence of the waves on the airflow is much more subtle and cannot be easily identified from instantaneous snapshots alone.

\begin{figure*}
\includegraphics[width=15cm]{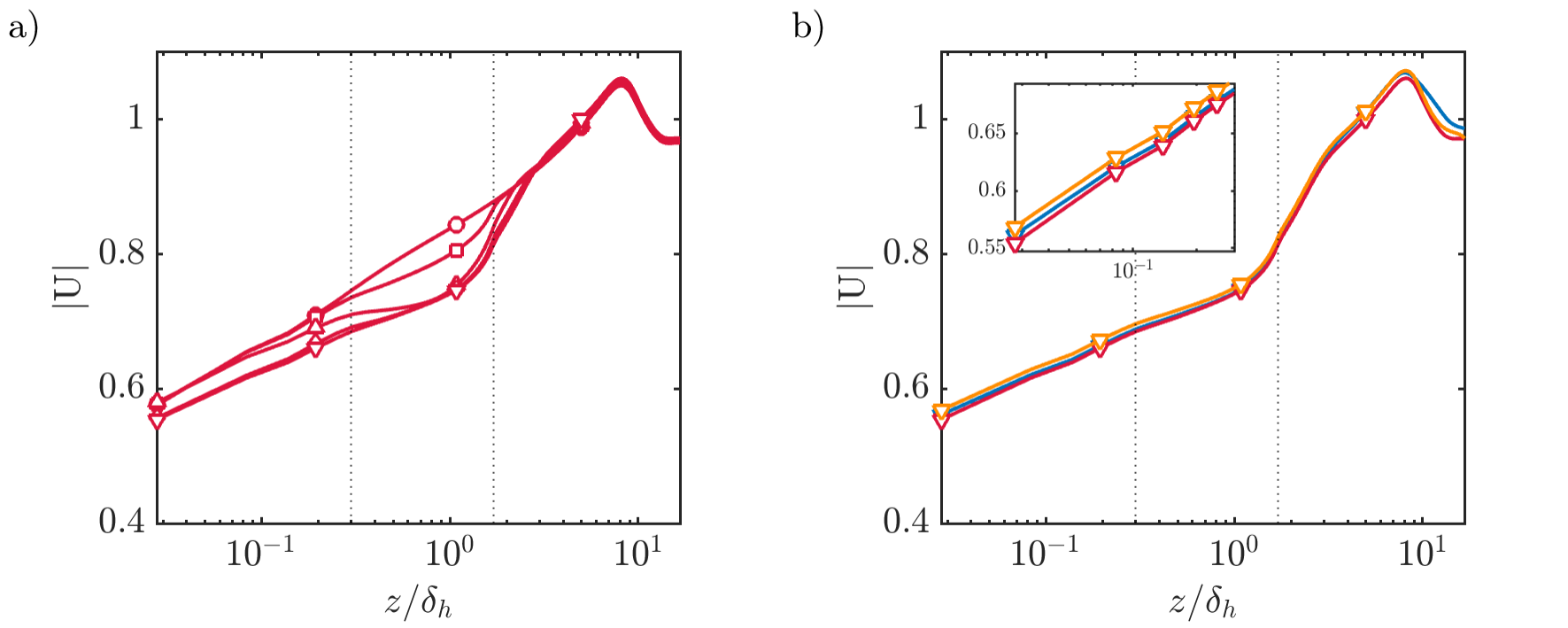}
\caption{\label{fig:fig9} Normalized vertical profile of the spanwise and time averaged wind velocity $|{\bf {U}}|=(u^2+v^2)^{1/2}$ at different streamwise locations. a) Case R3, where each symbol represents a distinct streamwise locations. Circle symbol: $x=-7D$ measured from the location of the first wind turbine row. Square symbol: $x=0D$ (location of the first turbine row). Upward-pointing triangle symbol: $x=7D$. Diamond symbol: $x=14D$ and Downward-pointing triangle symbol: $x=21D$. b) Comparison between R1 (blue), R2 (yellow) and R3 (red) case at $x=21D$. Inset shows a zoomed-in view below the turbine. For both figures, the vertical dotted line represents the bottom and top of the turbine-rotor planes. }
\end{figure*}

A clearer picture emerges from the mean velocity statistics. Figure~\ref{fig:fig9}a shows the mean horizontal wind-velocity profiles as function of height at several streamwise locations. The figure illustrates how the profile is progressively altered as the flow passes through successive turbine rows, and it also indicates that the wind-farm boundary layer becomes nearly fully developed by the third row. To isolate the effect of the waves, Fig.~\ref{fig:fig9}b compares the mean velocity profiles for the no-wave case (R1), the sinusoidal monochromatic-wave case (R2), and the broadband-wave case (R3). The differences are small, but systematic. Relative to the no-wave case, the broadband-wave case exhibits a slight downward shift of the velocity profile, while the monochromatic-wave case shows a slight upward shift, most noticeably below hub height.

The wave-induced differences observed in cases R2 and R3 follow the trend expected from the MOSD formulation, reflecting its prescribed dependence on the relevant wave parameters. In the form-drag contribution, the wave-induced stress depends on the relative velocity between the airflow and the moving surface, $(u-c)$, where in the present formulation $u=u_\Delta$. In these simulations, we obtain $c_p/u_\Delta \approx 0.95$ for the broadband-wave case (R3) and $c_p/u_\Delta \approx 1.1$ for the sinusoidal monochromatic-wave case (R2). Thus, in case R2 the wave travels slightly faster than the incoming airflow, leading to a small positive momentum flux from the wave to the air and, consequently, to the slight upward shift of the velocity profile. In contrast, for case R3 the peak wave travels slightly more slowly than the incoming airflow. Although both wave cases share the same peak-wave speed, the broadband-wave field also contains shorter resolved waves that travel more slowly than the peak wave. These slower wave components generate a somewhat larger relative drag, which is consistent with the slight downward shift of the velocity profile relative to the no-wave case. 

\subsection{\label{sec:power}Turbine Power Fluctuations}

We now turn to the central question of whether large-scale ocean waves can induce flow changes of sufficient magnitude to affect turbine power production. To address this question, we evaluate the power extracted by each turbine based on the turbine-induced force and the local wind velocity. Following \citet{calaf2010large}, the power of an individual wind turbine is computed as
\begin{equation}
    P_{ij} = \frac{1}{2}C_T'\frac{\pi}{4}D^2\langle u \rangle_D^3,
\end{equation}
where the subscript ``$ij$'' denotes the turbine located at the $i$th column and $j$th row, and $\langle u \rangle_d$ is the disk-averaged velocity.

To better identify possible wave-induced signatures in the turbine response, we analyze the time series of the aggregate wind-farm power over 160,000 consecutive time steps and perform a spectral analysis of these data. This time series represents approximately 1 hour of data with a $\Delta t=0.0234$ seconds. The time series is partitioned into 32 overlapping segments with 75\% overlap, each containing 32,768 samples. Because the signal is not periodic in time, a normalized Hanning window is applied to each segment to reduce spectral leakage. This procedure allows us to examine whether the imposed wave motion, as well as the smaller-scale waves present in the broadband case (R3), leaves a detectable imprint on the power fluctuations. The frequency resolution is chosen such that the smallest resolved frequency increment satisfies $\Delta\omega = \omega_p/100$, ensuring that the spectral analysis can resolve the peak-wave frequency and its possible influence on the turbine-power signal.

\begin{figure*}
\includegraphics[width=15cm]{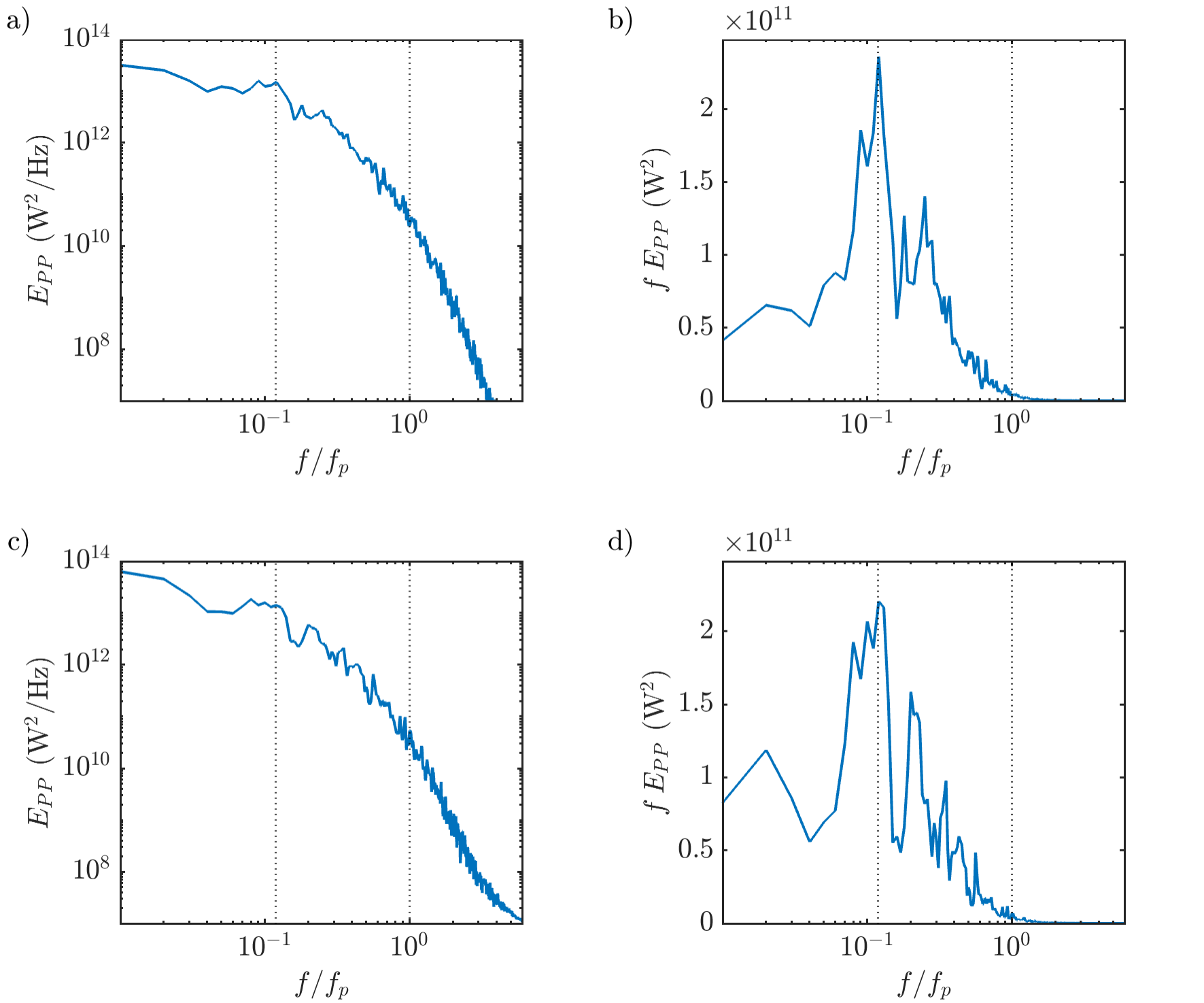}
\caption{\label{fig:fig10} (a-c) Power spectral density (PSD) of the total aggregate power of the offshore wind farm. (b-d)) Pre-multipied PSD of the total aggregate power. The first row is for the R1 case and the second row is for the R2 case. }
\end{figure*}

Figure~\ref{fig:fig10} shows the power spectral density of the total aggregated power for cases R1 and R2. In both cases, a pronounced peak is observed at approximately $f/f_p \approx 0.12$, particularly in the premultiplied spectra. This frequency corresponds to the passage of flow structures between successive turbine rows. From the WMLES, we estimate a mean advection velocity between rows of approximately $U \approx 13.5$ m/s. Using the streamwise turbine spacing, this gives $U/(S_x f_p) \approx 0.12$, confirming that the observed peak is associated with the temporal correlation induced by the downstream advection of turbine wakes from one row to the next. Similar behavior has been reported in previous wind-farm studies \citep{Bossuyt2017,Stevens2014Temporal}.

By contrast, no distinct wave-induced signature appears in the power spectra of case R2 at, or near, the imposed wave frequency, suggesting that the direct influence of the waves on the aggregate power signal is weak. This is even more apparent in the premultiplied spectra. The broadband-wave case (R3) shows a similar behavior, with no identifiable signature at either the peak-wave frequency or the frequencies of the smaller resolved waves, and is therefore omitted for brevity.

Although the PSD of the total power does not exhibit a clear wave-induced signature, this does not necessarily imply that the waves have no influence on the turbulent airflow within the turbine rotor region. Indeed, it has been argued that, under suitable simplifications, turbine power fluctuations are approximately linearly related to disk-averaged velocity fluctuations \citep{Bossuyt2017}, such that
\begin{equation}
    P'_i(t) \approx C_2\, \langle u_i'\rangle_D,
\end{equation}
where $C_2=(3/2)\rho A\,C_p \overline{\langle u \rangle}_D^2$, $\overline{\langle u \rangle}_d$ is the temporally and disk-averaged velocity, and $A$ is the rotor-disk area.

This relation suggests that if the waves have an effect on the flow velocity fluctuations at hub-height, some signature could be present in the power fluctuations as well. We therefore examine the dynamics of the velocity fluctuations at different vertical locations across the rotor plane in order to assess whether the waves leave a detectable imprint on the local turbulent flow. In this way, we can evaluate whether wave-induced changes in the rotor-region turbulence may contribute, even indirectly, to turbine power fluctuations.

To examine the spatial and temporal dynamics of the turbulent fluctuations at different vertical locations, we again employ spectral analysis by computing the three-dimensional Fourier transform of the streamwise velocity fluctuations. The resulting wavenumber-frequency spectrum is defined as
\begin{equation}
    E_{\mathrm{uu}}(k_x,k_y,\omega;z) =
    \frac{\overline{\hat{u}'(k_x,k_y,\omega;z)\,\hat{u}'^{\,*}(k_x,k_y,\omega;z)}}
    {\Delta k_x\,\Delta k_y\,\Delta \omega},
\end{equation}
where $\Delta k_x = 2\pi/L_x$ and $\Delta k_y = 2\pi/L_y$ are the streamwise and spanwise wavenumber resolutions, respectively. The overbar denotes averaging over all overlapping intervals, and $(\cdot)^*$ denotes complex conjugation.

To estimate these spatiotemporal spectra, we collect 160,000 consecutive snapshots of the streamwise velocity fluctuations in the precursor domain at several vertical heights, using a sampling interval of $\Delta t = 0.0234$ s. Because the LES employs a concurrent-precursor methodology, a normalized Tukey window is applied in the streamwise direction. Since the signal is not periodic in time, a normalized Hanning window is used. To improve statistical convergence, the data are partitioned into 32 overlapping intervals with 75\% overlap, each containing 32,768 samples.

Because the waves propagate in the same direction as the wind, that is, in the positive $x$--direction, we focus on the reduced $k_x$--$\omega$ spectrum obtained by integrating the two-dimensional horizontal wavenumber spectrum over the spanwise wavenumber,
\begin{equation}
    E_{\mathrm{uu}}(k_x,\omega;z) = \int E_{\mathrm{uu}}(k_x,k_y,\omega;z)\,dk_y.
\end{equation}

\begin{figure*}
\includegraphics[width=16cm]{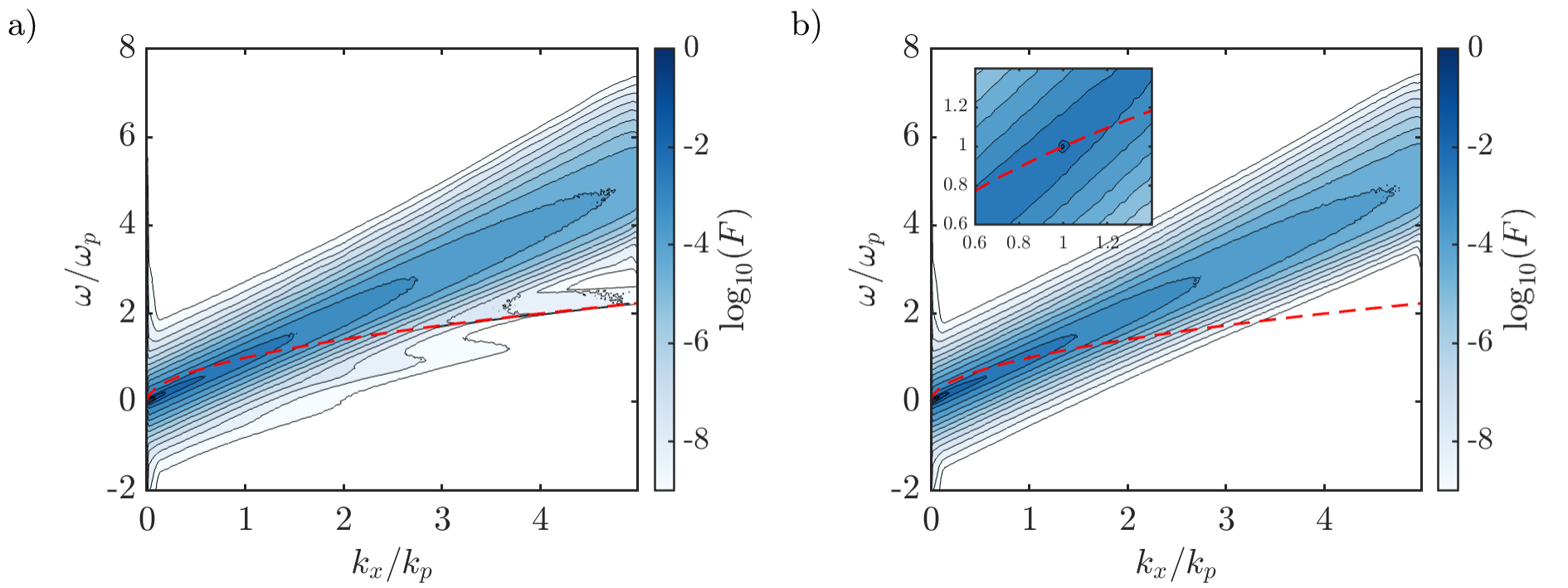}
\caption{\label{fig:fig11} Contours of the wavenumber-frequency spectrum for the streamwise velocity fluctuations $F = E_{\text{uu}} (k_x,\omega)/\max[E_{\text{uu}} (k_x,\omega)]$ for the a) R3 case and b) R2 case at a height $z= 3$ m. The red dashed line represents the normalized dispersion relation $\omega = \sqrt{g\,k}$. }
\end{figure*}

Figure~\ref{fig:fig11} shows the wavenumber--frequency spectrum of the streamwise velocity fluctuations at $z=3$ m for cases R2 and R3. In both cases, the spectra are tilted toward positive frequencies and exhibit Doppler broadening, a signature of random sweeping, whereby large-scale motions decorrelate the smaller turbulent scales in both space and time. Similar behavior has been reported previously in LES of canonical channel flows \citep{Wilczek_Stevens_Meneveau_2015,CHUNG_McKEON_2010}.

In addition to this broadband turbulent signature, both wave-forced cases exhibit a clear wave-induced imprint in wavenumber-frequency space. For the broadband-wave case (R3), the spectrum shows a distinct signature aligned with the dispersion relation of the surface waves. A similar dispersion-relation signature has also been observed in wave-resolving simulations \citep{hao_specwave_2019}. For the monochromatic-wave case (R2), the wave-induced signature appears as a single localized peak at the imposed wave frequency and wavenumber. This is consistent with the single-mode nature of the prescribed wave field. Moreover, the fact that this peak lies on the expected dispersion relation confirms that it is indeed associated with the surface wave.

Although these spectra are evaluated below the rotor region, they nevertheless demonstrate that the wave field leaves a clear imprint on the dynamics of the turbulent fluctuations. They also highlight the capability of the MOSD model to capture these wave-induced spatiotemporal signatures without explicitly resolving the moving surface.

\begin{figure*}
\includegraphics[width=15cm]{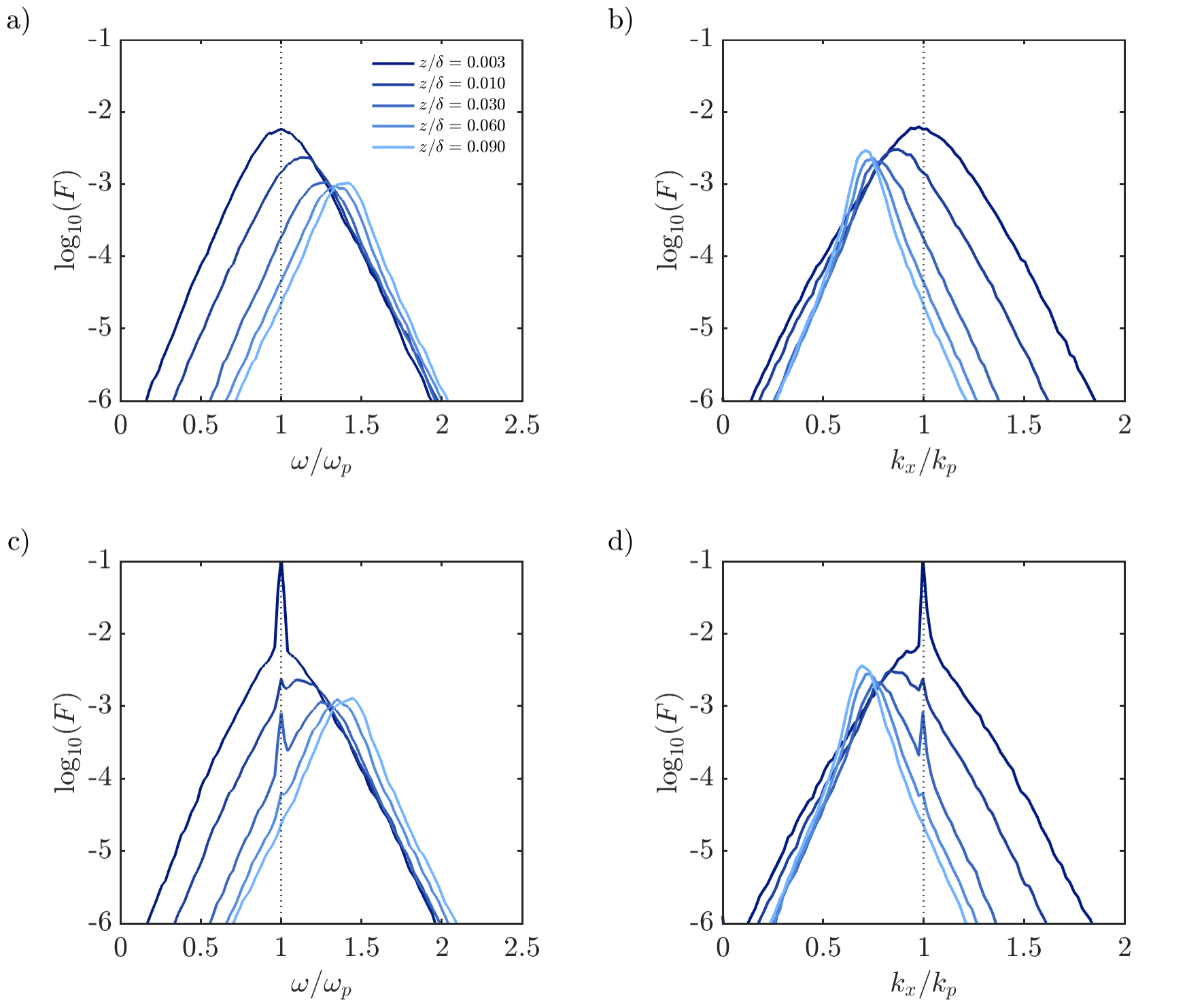}
\caption{\label{fig:fig12} Cuts through wavenumber-frequency spectra at several vertical heights for R1 case (first row, a,b, no wave case) and R2 case (second row, c,d, monocromatic wave case), at $k_x/k_p=1$ (a-c) and at $\omega/\omega_p=1$ (c-d). The dark blue to light blue colors represent going from low heights ($z/\delta =0.003$) to high ($z/\delta =0.09$), where $\delta=1000$ m. }
\end{figure*}

To better understand how the wave-induced signatures evolve with height, especially within the rotor region, Fig.~\ref{fig:fig12} shows cuts through the wavenumber-frequency spectra for the no-wave case (R1) and the monochromatic-wave case (R2). The main wave-induced feature is a localized peak at the imposed wave wavenumber and frequency. The results also show that this peak weakens progressively with increasing height in the boundary layer. This behavior is consistent with previous wave-resolving simulations \citep{hao_specwave_2019}, and more recent work has shown that the decay of such wave-induced peaks approximately follows  an exponential law of the form $\exp(-2k_p z)$ \citep{Ayala2026_madrid}, consistent with potential flow solution of flow over harmonic wave fields \citep{Kundu}.

In the present results, the wave signature remains visible throughout the lower portions of the rotor region, roughly between 30 and 60 m, but becomes negligible by hub height. By contrast, the broadband-wave case (R3) does not exhibit a similarly distinct peak at the frequency and wavenumber of the peak wave and is therefore not shown. It is not surprising that a wavefield with more distributed energy content would not leave a distinct single-frequency imprint on the flow.

This analysis suggests that, although the waves do leave a measurable imprint on the turbulent fluctuations within the rotor region, that imprint is relatively weak even for waves whose wavelength is comparable to hub height. It is therefore unlikely that these wave-induced fluctuations strongly affect turbine power fluctuations directly. This interpretation is further supported by the fact that power fluctuations depend on disk-averaged velocity fluctuations. As a result, even if the wave signature is noticeable in the lower portion of the rotor, its contribution may be substantially reduced once the velocity field is averaged over the full rotor disk, particularly since the upper portion of the rotor exhibits no wave-induced modulation.

\section{Conclusions}

In this study, we applied and tested the moving surface drag (MOSD) framework beyond the  original validation studies for monochromatic or simple waves \cite{ayala2024} and assessed its performance for increasingly complex applications, including Stokes-like waves, broadband wave spectra, offshore wind-farm boundary layers, and conventionally neutral offshore wind-farm flows over monochromatic waves.

For Stokes-like waves, the MOSD-based WMLES results were compared to experimental data and were shown to reproduced the measured mean streamwise velocity profiles quite well across the experimental cases considered. The model also captured the expected increase in drag relative to sinusoidal waves with the same primary wavelength and amplitude, consistent with the sharper crests and larger local slopes of the Stokes-wave geometry. Phase-averaged analysis showed that the model captures the qualitative structure of the streamwise wave-induced motions above the LES wall model matching height, although their amplitude was underpredicted relative to the experiments. This limitation is likely associated with the fact that the MOSD wall-stress formulation does not explicitly impose vertical wave-induced motions.

For broadband wave fields, the MOSD model was tested against wave-resolving LES data for JONSWAP wave spectra. The resulting mean velocity profiles agreed well with the reference simulations for all wave-age cases and for both coarse and fine grid resolutions. These results indicate that the model can represent the mean effects of broadband resolved waves while remaining robust to changes in horizontal wave-field resolution. They also show that unresolved wave components must be accounted for to provide the associated additional momentum fluxes. In this context, the SWARL model provided an effective and tractable estimate of the equivalent roughness length associated with the unresolved portion of the wave field.

The model was then applied to offshore wind-farm boundary layers over broadband waves and compared with wave-resolved LES of fully developed offshore wind farms. Across the wind-farm spacings considered, the MOSD-based WMLES reproduced the mean velocity profiles below, within, and above the turbine region without explicitly resolving the moving sea surface or using a wave-following coordinate transformation. These results show that the MOSD framework can capture the main mean-flow effects of broadband waves in wind-farm boundary layers while substantially reducing the modeling complexity associated with phase-resolved wave simulations.

Finally, exploratory simulations were performed for a fixed-bottom offshore wind farm under conventionally neutral atmospheric conditions including large-wavelength waves. The mean velocity profiles showed small but systematic wave-induced shifts. Spectral analysis of the aggregate farm power did not reveal a clear peak at the imposed wave frequency, indicating that the direct influence of these waves on total power fluctuations is very weak. However, wavenumber--frequency spectra of the streamwise velocity showed clear wave-induced signatures near the wave surface and, for the monochromatic case, detectable signatures into the lower part of the rotor region. These signatures decayed with height and became negligible near hub height, suggesting that they are largely filtered out when turbine power is estimated from the disk-averaged velocity.

Overall, the results show that MOSD model provides a computationally tractable framework for incorporating resolved wave effects into simulations of wind over waves and offshore wind farms. The model captures the main influence of moving ocean waves on mean momentum transfer and reproduces mean flow statistics from experiments and wave-resolved simulations across a range of wave configurations. Future work should focus on improving the representation of wave-coherent motions above the surface and extending the model so that it can accurately capture swell-wave regimes, where the relative motion between the waves and the airflow can lead to momentum transfer from the ocean to the atmosphere.

\begin{acknowledgments}
This work was supported by the National Science Foundation (via NSF grants CBET-2401013 and CBET-2401014) and the Department of Energy.
\end{acknowledgments}

\section*{Data Availability Statement}
The data that support the findings of this study are available from the corresponding author upon reasonable request

\appendix
\section{\label{lesgo_app}Large-Eddy Simulation Solver (LESGO)}

The MOSD model is implemented in the research LES code LESGO, originally developed following the work of \cite{albertson}. The open-source version of LESGO is available on GitHub \citep{lesgo} and includes several subgrid-scale stress parameterizations, wall models, and wind-turbine representations based on actuator disk and actuator line approaches. The code can be run with fully periodic boundary conditions or with the concurrent-precursor inflow-generation technique developed by \cite{stevens2014}. LESGO has been validated and widely used in studies of atmospheric boundary layers \citep{BouZeid,Narasimhan2024} and wind-energy applications \citep{ghanesh2022,calaf2010large,stevens2014}. In this section, we briefly summarize the governing equations and numerical methods used in the solver.

LESGO solves the filtered Navier--Stokes equations in rotational form, together with the scalar potential-temperature transport equation when buoyancy effects are included. The use of the rotational form helps ensure mass and kinetic-energy conservation. The governing equations are
\begin{equation}
\frac{\partial u_i}{\partial x_i} = 0,
\end{equation}

\begin{widetext}
\begin{equation}
\frac{\partial u_i}{\partial t} + u_j \left(\frac{\partial u_i}{\partial x_j} - \frac{\partial u_j}{\partial x_i}\right)
= - \frac{1}{\rho_0}\frac{\partial P_{\infty}}{\partial x_1}
- \frac{\partial P}{\partial x_i}
- \frac{\partial \tau_{ij}}{\partial x_j}
+ \nu \frac{\partial^2 u_i}{\partial x_j^2}
+ \frac{g}{\theta_0}(\theta-\theta_0)\delta_{i3}
+ \frac{1}{\rho_0}f_x\delta_{i1}
+ \frac{1}{\rho_0}f_y\delta_{i2}
- f_c u \delta_{i2}
+ f_c v \delta_{i1},
\label{eq:momentum}
\end{equation}
\end{widetext}

\begin{equation}
\frac{\partial \theta}{\partial t} + u_j \frac{\partial \theta}{\partial x_j}= \frac{\partial \Pi_j}{\partial x_j},
\label{eq:scalar}
\end{equation}
where $u_i=(u,v,w)$ for $i=1,2,3$ are the filtered velocity components in the streamwise ($x$), spanwise ($y$), and vertical ($z$) directions, respectively.

In Eq.~\ref{eq:momentum}, $\tau_{ij}=\sigma_{ij}-\frac{1}{3}\sigma_{kk}\delta_{ij}$ is the deviatoric part of the subgrid-scale (SGS) stress tensor $\sigma_{ij}$. The quantity $P=P_*/\rho_0+\frac{1}{3}\sigma_{kk}+\frac{1}{2}u_i^2$ is the modified pressure, where the actual pressure $P_*$ divided by the reference density $\rho_0$ is augmented by the trace of the SGS stress tensor and the kinematic pressure associated with writing the nonlinear terms in rotational form. The deviatoric SGS stress is modeled using an eddy-viscosity closure,
\begin{equation}
\tau_{ij} = -2\nu_T S_{ij},
\end{equation}
where $\nu_T$ is the SGS eddy viscosity and $S_{ij}$ is the resolved strain-rate tensor. The eddy viscosity is given by
\begin{equation}
\nu_T = (C_S \Delta)^2 \sqrt{S_{ij}S_{ij}},
\end{equation}
where $C_S$ is the Smagorinsky coefficient and $\Delta=(\Delta x \Delta y \Delta z)^{1/3}$ is the filter scale based on the local grid spacings. In the present study, $C_S$ is determined dynamically using the Lagrangian scale-dependent model of \citep{BouZeid}.

The forcing terms $f_i=(f_x,f_y,0)$ represent the streamwise and spanwise components of the turbine force exerted on the fluid. The effect of the wind turbine rotor on the wind is modeled using the actuator-disk model (ADM), which is commonly applied to the study of land-based wind farms \citep{Jimenez2010WakeDeflection,calaf2010large,shapiro2019} and offshore wind farms \citep{yang2014large}. The term $-(1/\rho_0)\partial P_\infty/\partial x_i$ denotes the externally imposed mean pressure gradient used to drive the flow. For isothermal and neutrally buoyant cases, such as those considered in Sects.~\ref{sec:stokes_waveind}, \ref{sec:spectrum}, and \ref{sec:owbl_broadband}, Eq.~\ref{eq:scalar} is not solved and the buoyancy term in Eq.~\ref{eq:momentum} vanishes.

For conventionally neutral atmospheric-boundary-layer cases, such as those considered in Sect.~\ref{sec:owbl_realistic}, the mean pressure gradient is written in terms of the geostrophic velocity components $U_g$ and $V_g$ through geostrophic balance,
\begin{equation}
\frac{1}{\rho_0}\frac{\partial P_\infty}{\partial x}=f_cV_g,
\qquad
\frac{1}{\rho_0}\frac{\partial P_\infty}{\partial y}=-f_cU_g,
\end{equation}
where $U_g=G\cos\alpha$ and $V_g=G\sin\alpha$, with $G=(U_g^2+V_g^2)^{1/2}$ the magnitude of the geostrophic wind and $\alpha$ the angle of the geostrophic wind relative to the streamwise direction. In the Coriolis terms, $f_c=2\Omega\sin\phi=10^{-4}\,\mathrm{s}^{-1}$ is the Coriolis parameter at latitude $\phi=45^\circ$. The Kronecker delta $\delta_{ij}$ in Eq.~\ref{eq:momentum} specifies the directions of the buoyancy, turbine, and Coriolis forcing terms. In the buoyancy term, $g=9.81\,\mathrm{m/s^2}$ is the gravitational acceleration and $\theta_0$ is the reference potential temperature.

In Eq.~\ref{eq:scalar}, $\Pi_j$ denotes the SGS heat flux. Because the equations are solved in the high-Reynolds-number limit, molecular viscosity and molecular heat diffusion are neglected in Eqs.~\ref{eq:momentum} and \ref{eq:scalar}. The required scalar diffusion is instead modeled through the SGS heat flux,
\begin{equation}
\Pi_j = \alpha_T \frac{\partial \theta}{\partial x_j},
\end{equation}
where $\alpha_T$ is the SGS heat diffusivity. The SGS momentum and heat diffusivities are related through the SGS Prandtl number, $Pr^{\rm SGS}=\nu_T/\alpha_T$, so that
\begin{equation}
\alpha_T = \frac{1}{Pr^{\rm SGS}}(C_S \Delta)^2 \sqrt{S_{ij}S_{ij}}.
\end{equation}

Spatial discretization is performed using a pseudo-spectral method in the streamwise and spanwise directions and a second-order centered finite-difference scheme in the wall-normal direction. Time advancement is carried out using a second-order Adams--Bashforth scheme.

\section{\label{gridmosd_app}Grid Sensitivity Analysis of the MOSD Model}

In this section, we examine the sensitivity of the MOSD model to grid resolution and to the choice of reference height, as discussed in Sect.~\ref{sec:delta_selec}. Specifically, we compare two definitions of the reference height used to evaluate the incoming velocity in the pressure-stress model of Eq.~\ref{eq:mosd_wpm}. In the first, the reference height is selected according to the geometry-based criterion proposed in Sect.~\ref{sec:delta_selec}, namely $\Delta = 3H_p$. In the second, the reference height is taken to depend directly on the horizontal grid resolution, $\Delta = \Delta x$, which is the choice adopted in our previous study \citep{ayala2024}.

To assess these two choices, we compare WMLES results against the experimental measurements of \citep{yousefi_2020}, who studied turbulent airflow over monochromatic waves. We select this configuration because it was also used in our previous validation of the MOSD model, allowing for a direct assessment of how the reference-height choice affects the predictions. In particular, we focus on the highest-wave-steepness case, which provides a stringent test of the model. The monochromatic wave has amplitude $a=0.0196$ m and wavelength $\lambda=0.48$ m, corresponding to a wave steepness of $ak\approx 0.26$. The boundary-layer height is $\delta=0.275$ m and the friction velocity is $u_*=0.567$ m/s. All of these wave and flow parameters are taken from \cite{yousefi_2020}.

\begin{figure*}
\includegraphics[width=15cm]{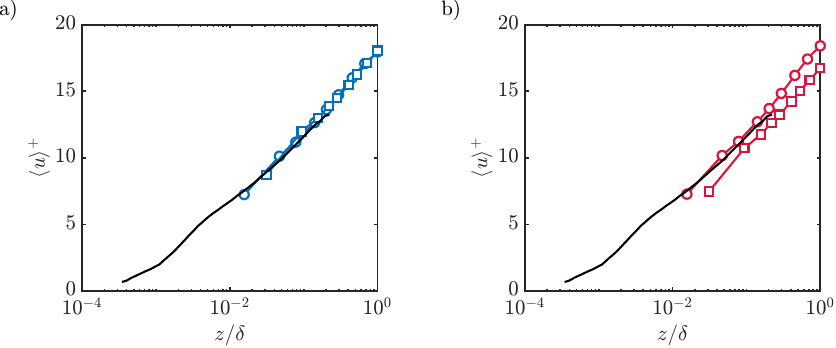}
\caption{\label{fig:app1} Mean streamwise velocity profiles for the $ak=0.26$ case from \cite{yousefi_2020}. a) WMLES using reference height $\Delta=3Hp$ and b) WMLES using reference height $\Delta=\Delta x$. Solid line: Experimental data from \cite{yousefi_2020}. Circles: Fine grid resolution. Square: Coarse grid resolution.}
\end{figure*}

The numerical domain is chosen as $L_x \times L_y \times L_z = 6\lambda \times 3\lambda \times \delta$. Two grid resolutions are considered: a fine grid with $N_x \times N_y \times N_z = 64 \times 32 \times 32$ and a coarse grid with $N_x \times N_y \times N_z = 32 \times 16 \times 16$. Figure~\ref{fig:app1} shows the mean streamwise velocity profiles obtained with the MOSD model for both reference-height definitions and both grid resolutions.

The results show that using the geometry-based reference height, $\Delta = 3H_p$, yields the best agreement with the experimental data and remains relatively insensitive to the horizontal grid resolution. By contrast, the grid-dependent choice, $\Delta = \Delta x$, gives good agreement only on the fine grid; on the coarse grid it overpredicts the drag, leading to an excessive downward shift of the velocity profile relative to the experiments. It is also worth noting that the results obtained with $\Delta=\Delta x$ on the fine grid are very similar to those obtained with $\Delta=3H_p$. This is because, for the fine-resolution case, the two length scales are nearly identical: $\Delta x = 0.045$ m, while $3H_p = 0.046$ m. This close agreement further supports the interpretation that the discrepancy on the coarse grid arises primarily from the grid dependence of $\Delta=\Delta x$, rather than from any intrinsic difference between the two formulations.

\section{\label{swarl_app}Surface Wave Aerodynamic Roughness Length (SWARL) Model}

The Surface Wave Aerodynamic Roughness Length (SWARL) model \cite{ayalaSWARL2026} provides a physics-based estimate of the equivalent aerodynamic roughness length induced by a moving wave surface. The model is closely related to the MOSD framework, since it is built from the same two contributions to the total drag: a pressure or form-drag contribution associated with the resolved wave geometry, and a viscous/unresolved contribution associated with tangential stress and subgrid roughness. In this sense, the SWARL model may be viewed as a roughness-length formulation derived from the same physical ingredients as the MOSD wall-stress model. The formulation given below follows \cite{ayalaSWARL2026}.

In the SWARL framework, the total drag is expressed in terms of a nondimensional parameter $\Lambda$, which represents the net drag induced by the moving wave surface. This parameter is obtained by averaging the pressure-drag and unresolved-drag contributions over the wave surface. The pressure contribution depends on the local wave slope, the local wave propagation velocity relative to the airflow, and the windward/leeside asymmetry used in the MOSD model. The unresolved contribution is modeled through the same equilibrium roughness closure used in MOSD, represented by a friction factor $C_f$.

For a given surface elevation field $\eta(x,y,t)$, the SWARL model defines $\Lambda$ implicitly through
\begin{widetext}
\begin{equation}
\Lambda = \left\langle 
\frac{\alpha}{\pi+\alpha}
\left\{
\left[(1-C_x^+\Lambda^{1/2})\hat{n}_x\right]^2
+
\left[C_y^+\Lambda^{1/2}\hat{n}_y\right]^2
\right\}
\frac{\partial \eta}{\partial x}
H\left[
(1-C_x^+\Lambda^{1/2})\hat{n}_x
-
C_y^+\Lambda^{1/2}\hat{n}_y
\right]
\right\rangle_{x,y}
+\frac{1}{2}C_f ,
\label{eq:Lambda_swarl}
\end{equation}
\end{widetext}
where $C_i^+=C_i/u_*$ are the local wave-velocity components normalized by the friction velocity, and $\langle\cdot\rangle_{x,y}$ denotes averaging over the horizontal surface. The friction factor is evaluated as $C_f=C_f(\Lambda^{-1/2}\Delta^+,z_0^u)$, where $\Delta^+=\Delta u_*/\nu$. Thus, $\Lambda$ depends on both the resolved wave geometry and the unresolved roughness contribution.

Once $\Lambda$ is obtained, the equivalent aerodynamic roughness length is evaluated as follows,
\begin{equation}
z_0 = \Delta \exp{\left(-\kappa \Lambda^{-1/2}\right)}.
\label{eq:z0_swarl_app}
\end{equation}
In the present work, we take $\Delta=3H_p$, where $H_p$ is the dominant positive surface height defined from the surface elevation field. When the SWARL model is applied only to the unresolved portion of the wave field, the resulting roughness length is interpreted as the unresolved roughness length $z_0^u$.

In practice, the SWARL model is evaluated numerically from a known surface realization $\eta(x,y,t)$, or from the filtered-out portion of the wave field when only unresolved waves are of interest. Because the model combines the same pressure-drag and viscous/unresolved components used in MOSD, it provides a natural way to convert unresolved wave geometry into an equivalent aerodynamic roughness length.

\bibliography{ayala_lit}

\end{document}